\documentclass[a4paper, 12pt]{iopart}   
\usepackage{iopams}
\usepackage{amsarrows}
\newcommand{\Cov}{\mathop{\mathrm{Cov}}\nolimits}
\newcommand{\Var}{\mathop{\mathrm{Var}}\nolimits}
\newcommand{\mean}{\mathop{\mathrm{E}}\nolimits}
\newcommand{\sinc}{\mathop{\mathrm{sinc}}\nolimits}
\newcommand{\sign}{\mathop{\mathrm{Sign}}\nolimits}
\renewcommand{\Re}{\mathop{\mathrm{Re}}\nolimits}

\usepackage[bookmarks]{hyperref}

\usepackage{graphicx}
\usepackage{subfig}
\usepackage[below]{placeins}

\begin{document}

\title[Nodal Intersections With a Reference Curve]
{The Statistics of the Points Where Nodal Lines Intersect a Reference Curve}

\author{Amit Aronovitch and Uzy Smilansky}

\address{Department of Physics of Complex Systems,\\
The Weizmann Institute of Science, 76100 Rehovot, Israel}

\ead{amit.aronovitch@weizmann.ac.il}

\begin{abstract}
 We study the intersection points of a fixed planar curve
$\Gamma$ with the nodal set of a translationally invariant and
isotropic Gaussian random field $\Psi(\bi{r})$ and the zeros of its
normal derivative across the curve.
The intersection points form a discrete random process which is the
object of this study. The field probability distribution function
is completely specified by the correlation 
$G(|\bi{r}-\bi{r}'|) = \left \langle\Psi(\bi{r}) \Psi(\bi{r}') \right \rangle$.
 Given an arbitrary 
$G(|\bi{r}-\bi{r}'|)$,  we compute the two point correlation
function of the point process on the line, and derive other
statistical measures (repulsion, rigidity) which characterize the
short and long range correlations of the intersection points. We
use these statistical measures to quantitatively  characterize the
complex patterns displayed by various kinds of nodal networks. We
apply these statistics in particular to nodal patterns of random
waves and of eigenfunctions of chaotic billiards.
Of special interest is the observation that for monochromatic random waves, the
number variance of the intersections with long straight segments grows like
$L \ln L$, as opposed to the linear growth predicted by the percolation model,
which was successfully used to predict other long range nodal properties of that
field.

\end{abstract}


\section{Introduction}
\label{sec:intro}
Entangled planar networks of convoluted lines appear quite often
in various studies in physics and mathematics. To cite a few
examples, recall the snapshots of polymer solutions, the
level sets of rugged terrain, the trajectories of
Brownian particles~\cite{johansson:path_rmt}, the domain boundaries 
in magnetic materials or in simulations of random percolation
\cite{stauffer:percolation} {\it etc}. To characterize such networks in a
concise and quantitative way, one usually recourses to statistical
measures, chosen to describe the specific properties of the
network which are relevant to the problem at hand.  Some studies
give bounds on the length or the curvature of the nodal lines of
wave functions in bounded domains, in other instances, the
distribution of fractal dimensions of Brownian trajectories which
provide an impression of their ruggedness~\cite{stauffer:mandelbrot}.
Other examples are the distribution of the areas of connected domains in
critical percolation which show a universal power law, or
the (properly normalized) number of bulk nodal domains in billiard
wave functions which follow universal patterns that distinguish
between chaotic and integrable billiards~\cite{blum:nd_stats}, 
\cite{bogomolny:percolation}. None of
these measures provide a complete description of the complexity of
the network under study, and there is always room for introducing
new measures which shed light on features which were not brought
to the front by previous studies.

In the present work we are interested in the point process
generated by the intersections  of a complex network of lines with
a given reference curve. To the best of our knowledge, previous studies
of the statistics of such point processes were not as detailed as the
corresponding analysis for other systems (such as energy spectra, eigenvalues
of random matrices, or one dimensional gas). In particular, while the density
and correlation function were calculated, other physically meaningful 
observables which can be extracted from them, such as the number variance,
rigidity and power spectrum were not investigated.

Longuet-Higgins, in his analysis of the moving
surface of the sea~\cite{longuethiggins}, derived the density and
correlation functions of the zeros on a straight line.
Berry and Dennis~\cite{berry:sing_rw} studied the distribution of phase
singularities in a random complex field. For a planar field, these 
singularities are actually the intersections of the nodal lines of the real
part of the field with those of the imaginary part.
In these cases (and others as well, including the present paper), restricting
the field to a straight line allows using the formula derived by 
Rice~\cite{rice:rnoise2} for calculating the exact correlation of the zeros
from the known two point correlation of the amplitudes.
Blum \etal~\cite{blum:nd_stats} considered 
billiard wave functions, and studied the number of intersections of 
their nodal lines with the billiard boundary. It was shown that the
distributions of the (properly normalized) number of intersections
distinguish between chaotic and integrable billiards. 
Johansson~\cite{johansson:path_rmt}
considered the trajectories of $N$ Brownian particles on a line,
which start equally separated at $t=0$ and return to their original
position at $t=2T$ without intersecting each other's path.
The distribution of spacings at $t=T$, measured in units of the average
spacing, was shown to be identical to the distribution of eigenvalues in the
Gaussian Unitary Ensembles of $N\times N$ matrices (GUE).
Baik and Rains~\cite{baik:niwalk_goe} showed that for a certain class of 
discrete, non-intersecting random walkers, when not restricted to return to the
original position, the limiting distribution of spacings matches that of the 
eigenvalues of the Gaussian Orthogonal ensemble (GOE). Such problems, when 
considered in $(x,t)$ space, represent intersections of random
planar paths with a reference line orthogonal to the $t$
axis.

In the present paper we shall study the points generated by the
intersection of a reference curve with the nodal lines of three fields.

\begin{enumerate}
\item \noindent Random monochromatic waves~\cite{berry:smc_wf}
in $\mathbb{R}^2$ which are solutions of the wave equation
\begin{equation}
-\Delta \Psi(x,y) = k^2 \Psi(x,y)
 \label{eq:waveeq}
\end{equation}
Since no boundary conditions are required, the solutions are plane
waves which propagate in arbitrary directions with wave vectors
$|\bi{k}|=k$.  The random wave ensemble can be constructed as
a linear superpositions of plane waves:
\begin{equation}
\label{eq:randomwave}
 \sqrt{\frac{2}{N}} \sum_{n=1}^N \cos ( \bi{k}_n \bi{r}+\phi_n)
\end{equation}
 (with $\bi{r} \in \mathbb{R}^2$, $|\bi{k}_n| = k$ and $N\gg 1$),
where $\bi{k}_n/|k|$ and $\phi_n$ are distributed uniformly 
and independently on the unit circle.
Equivalently, one can also use random
superpositions of solutions of the wave equation (\ref{eq:waveeq})
in polar coordinates
\begin{equation}
a_0 J_0(kr)+2\sum_{l>0} a_l J_l(kr)\cos(l\theta +\phi_l)
\label{eq:bessel}
\end{equation}
with real coefficients $a_l$, which are identically and independently
distributed Gaussian variables, and where the phases $\phi_l$ are
independent and uniformly distributed on $[0,2\pi]$.

The correlation function for this Gaussian ensemble is
\begin{equation}
G_{\mathrm{RW}}(\bi{r},\bi{r}') = J_0(k|\bi{r}-\bi{r}'|)
 \label{eq:corfunrw}
\end{equation}

\item \noindent Given a reference curve, we consider the normal derivative 
$\bi{n}\cdot\bi{\triangledown}\psi(\bi{r})$ of the random wave 
field~\eref{eq:randomwave}, where $\bi{n}$ is the direction normal to
the reference curve at $\bi{r}$.
For example, when the curve is the $x$ axis, the two point correlation of this
function is
\begin{equation}
G_\mathrm{NRW}(x,x') = 2 J_1(k|x-x'|)/(k|x-x'|)
\label{eq:corfunndrw}
\end{equation}
The zeros of the normal derivative are not, strictly speaking, nodal 
intersections, 
but they do define a point process, which can be studied the same way.
Furthermore, when the reference curve is straight,
these zeros are the nodal intersections of an actual random field.

The boundary modified random waves~\cite{berry:nodalstat}, were
introduced in order to study the effect of a Dirichlet boundary on
the statistics of the nodal set. 
They are defined in the upper half plane and are solutions 
of~\eref{eq:waveeq} subject to the boundary condition
\begin{equation}
\Psi(x,y=0) = 0
\label{eq:xaxis_boundary}
\end{equation}

As explained in~\sref{sec:nderiv}, the nodal intersections of that field with
its boundary, the $x$ axis, are identical with the zeros of the normal 
derivative of a Gaussian random wave field on this line. 
This fact makes the normally derived random wave field on a curved line a
natural candidate for comparison with generalized boundary modified models,
such as Wheeler's generalization for a circular 
boundary~\cite{wheeler:curv_bnd}, or the semi-classical treatment of a general
boundary by Urbina and Richter~\cite{urbina:sc_rw}.

\item \noindent Random Gaussian fields with the short range correlation function
\begin{equation}
G_{\mathrm{SRF}}(\bi{r},\bi{r}') = \exp(-k^2 {|\bi{r}-\bi{r}'|}^2/4)
 \label{eq:corfunsrf}
\end{equation}
were introduced in~\cite{foltin:linedist} to investigate the importance
of the fact that the random waves correlation function decays very
slowly. The correlation function above coincides with $G_{\mathrm{RW}}$ at
short ranges, but decays quickly at large values.
\end{enumerate}

According to the Uhlenbeck  theorem~\cite{uhlenbeck:eigenfunc}, the
crossing probability of nodal lines of random waves vanishes. 
In~\cite{monastra:avoided}, the ``avoidance'' of nodal lines was defined,
as a measure of the distance between nodal lines which avoid intersection
in the vicinity of saddle points. The probability distribution
$P(d)$ of the avoidances $d$  was computed and it displayed linear
repulsion at small values of $d$, $P(d) \propto d$. However the
coefficient of proportionality was not the same as one would get
from the application of Johansson's result to the case of Brownian
trajectories.

Having in mind the results of Johansson's paper on the one hand,
and knowing that nodal lines avoidances repel linearly on the
other hand, one might expect that the statistics of nodal intersections
would also follow (at least to some extent) the predictions of
random matrix theory. We shall show below that this is not the
case.

This paper is constructed as follows. In \sref{sec:zer_stat}, we discuss the
derivation of the nodal intersection 
statistics from the known distribution of the amplitudes. First, we extract
the necessary one-dimensional correlation functions which follow from
restricting the field
to a general reference curve (similar to the restriction of the amplitude to a
straight line, which was done in~\cite{longuethiggins}).
Then, the statistics of the zeros is extracted from these one-dimensional 
amplitude correlations. An exact formula for computing zeros correlations from
one dimensional amplitude correlations was derived by Rice in his analysis of 
random noise~\cite{rice:rnoise1},\cite{rice:rnoise2}. The derivation described
here, uses a different method, and is similar to the derivation used by other
authors~\cite{bogomolny:pol_zeros},\cite{berry:sing_rw}. However, as opposed
to these studies, we consider the case of a general curve, and derive
the formulae for the line and the circle as special cases. Restriction to a
general curve does not preserve the field's translational invariance. This
makes the results more complicated, but the methods remain the same. Such 
a generalization is necessary in cases where the test curve is inherent in the
problem we wish to model (such as the boundary of a quantum billiard), and 
not independent of the field. In~\ref{sec:curv_ndrv_ampcorr} we describe
the generalized calculation for the normally derived random waves.

In \sref{sec:mrw}, we apply the above to the monochromatic random wave 
\eref{eq:randomwave}, using a circle and a straight line for reference curves.
A universal correlation function is extracted for the semiclassical ($k\gg 1$)
limit, and found to decay slower than the RMT correlations. The number variance
and power spectrum are calculated and their special features, which are clearly 
different from the RMT predictions are shown as well. Over large ranges, the
number variance is shown to grow as $L \ln L$. This is in contrast with the
linear growth we might expect from Bogomolny's percolation 
model~\cite{bogomolny:percolation}, whose predictions for other asymptotic
properties of the field (nodal count~\cite{blum:nd_stats} and SLE driving 
force~\cite{bogomolny:sle}) give a satisfactory match to numerical data.
In \sref{sec:nderiv} we discuss the zeros distribution of the normal derivative
 of a Gaussian field on
the reference curve. In the framework described above this may be considered
a different form of restricting the field to the curve. The semiclassical limit,
number variance and form factor are calculated for this case too.

Finally, in~\sref{sec:other} we compare the above results to those of other
fields. The statistics mentioned above are calculated for the short-range
Gaussian model~\eref{eq:corfunsrf} (which displays a different behaviour),
and also evaluated numerically for chaotic billiards, which show a reasonable
match to the monochromatic random wave model. The theoretical properties of
the distributions considered are summarized in~\tref{cap:comparetab}, and
compared to the well known results for the Poisson process and RMT ensembles.

\section{Statistics of zeros from amplitude correlations }
\label{sec:zer_stat}

\subsection{Nearest neighbour distribution}
\label{sub:nn_dist}
\begin{figure}[htbp]
\centering
\includegraphics[clip,scale=0.8]{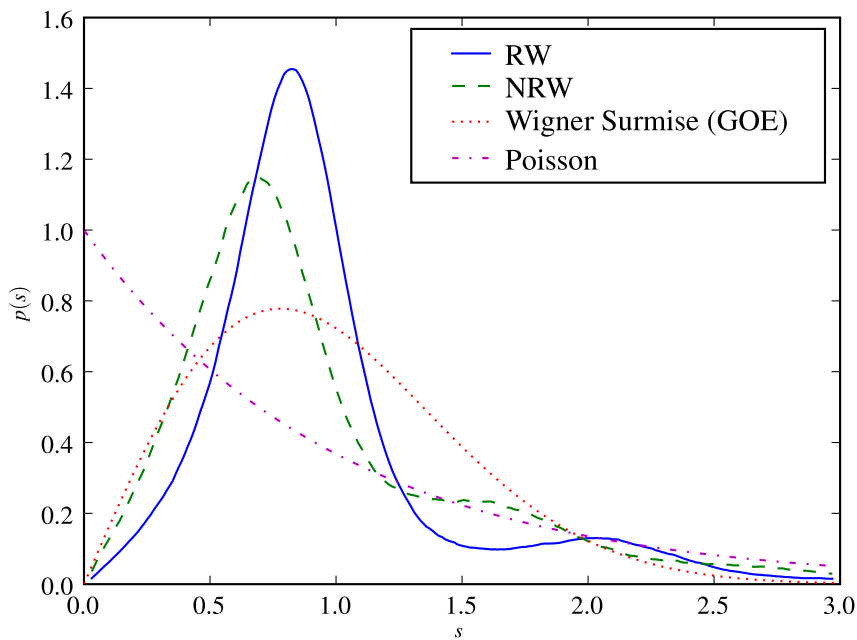}
\caption{\label{cap:rw_drw_nearest}Density of the nearest neighbour level 
spacing.}
\end{figure}

When considering the statistics of a sequence of points, the nearest neighbour
spacing distribution (where the spacing is measured in units of the mean 
spacing) is perhaps the most natural, and easy to evaluate experimentally. 
In~\fref{cap:rw_drw_nearest} we show the nearest neighbour distributions for two
point processes. The first is the intersections of the nodal lines of the random
wave (RW) field~\eref{eq:randomwave} with the reference curve.
The second process (NRW) is the zeros of the normal derivative of the
random waves on the curve. The distributions were generated by numerical
simulations, and compared with the corresponding statistics of the
random (Poisson) ensemble and the Wigner surmise (corresponding to the GOE
ensemble)~\cite{mehta:rand_mat}.

\begin{figure}[htbp]
\centering
\includegraphics[clip,scale=0.8]{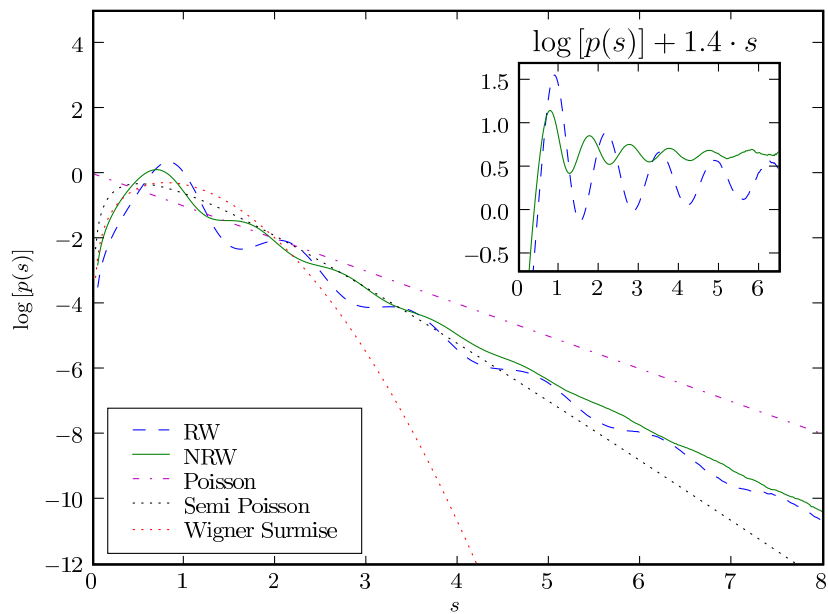}
\caption{Nearest neighbour level spacing (log density).
Inset shows persistent oscillations relative to the mean decaying curve.}
\label{cap:rw_nearest_big}
\end{figure}

At short ranges, both ensembles display linear level repulsion similar to the 
GOE distribution, but with different slopes.
The most conspicuous differences appear at large spacings, and to get a clearer
impression, we show in~\fref{cap:rw_nearest_big}, the same data in semi-log plot.
We observe two important differences.
\begin{enumerate}
\item \noindent On average, both distributions decay exponentially with
approximately the same rate $p(s)\sim \exp(-1.4 s)$. This decay is faster than the 
$\exp(-s)$ Poisson decay, but slower the ``Semi Poisson''~\cite{bogomolny:instat} 
distribution, which decays like $\exp(-2 s)$.
\item \noindent The overall exponential decays are decorated by persistent 
oscillations, which are clearly seen in the inset of~\fref{cap:rw_nearest_big}.
The oscillations have slowly decaying amplitudes and
their frequencies are $\sim 1.4\pi$ for RW and $\sim 2\pi$ for NRW\@.
\end{enumerate}

The results above demonstrate that the nodal intersection statistics are
significantly different than those of other, well known, point processes.
The purpose of this work is to investigate the reasons for the occurrence of 
these differences.
However, the nearest neighbour statistic is not easily amenable to analytic 
derivation. Rather, we use the two point correlation function, which is readily
accessible, as will be shown in this paper. For the asymptotic $s\ll 1$ limit,
this statistic coincides with the nearest neighbour density. Furthermore,
other statistics with well defined physical meaning, such as the power spectrum
and number variance can be expressed in terms of the correlation. We will now
give some definitions, and follow with an overview of a derivation for the
two point correlation of nodal intersections with a general reference curve.

\subsection{Two point correlations}
A random field $\psi(\bi{r})$ is Gaussian if for any vector $\bi{u}$ whose
elements $u_1,u_2,\ldots,u_n$ are field amplitudes $u_i=\psi(\bi{r}_i)$ or
partial derivatives of the field at some points $\bi{r}_1,\ldots,\bi{r}_n$,
the probability distribution of $\bi{u}$ is multivariate normal. The 
probability density of a multivariate normal random vector $\bi{u}$, with mean
value $\bi{u}_0$ is given by
\begin{equation}
\fl p(\bi{u})=\frac{1}{{(2\pi)}^{n/2}\sqrt{\det C}}
          \exp\!\!\left[ -\frac{1}{2}
                         {(\bi{u}-\bi{u}_0)}^T C^{-1} 
                         (\bi{u}-\bi{u}_0) \right],
\label{eq:multivar_denst}
\end{equation}
where $C$ is the covariance matrix $C_{i,j}=\Cov(u_i,u_j)$. For the random
wave ensemble~\eref{eq:randomwave}, Gaussianity follows from the $n$
dimensional central limit theorem.

Assuming smoothness of the field, means and covariances of the field's
derivatives can be calculated by deriving the appropriate amplitude statistics.
Combining that with the fact that all the elements of $\bi{u}_0$ and $C$ are
single point means and two point covariances, it follows that all
statistical properties of a Gaussian field $\psi(\bi{r})$ are completely
determined by its mean value $\langle \psi(\bi{r}) \rangle$ (which will be 
zero for the relevant cases investigated) and the two point covariance 
function 
$G(\bi{r},\bi{r'}) \equiv \Cov(\psi(\bi{r}),\psi(\bi{r}'))$. 
As a consequence, we should
be able to extract the distribution of the zeros from this function,
and, in particular, the correlation function of the nodal intersections can
be derived from the correlation function of the 1 dimensional restriction 
$f(t) \equiv \psi(\bi{r}(t))$
of the investigated field to the given reference curve 
$\Gamma:[0,L]\rightarrow \mathbb{R}^2 \equiv \bi{r}(t)$ ($t$ is the natural 
curve parameter).

The density and correlations of nodal intersections (NI) can be calculated directly
once $f(t)$ and its properties are known, in particular we shall use the two point
correlations of the restricted field $f$ and its derivative along the curve
$\dot{f}(t)\equiv \rmd f / \rmd t$:
\begin{eqnarray}
\label{eq:corrdefs}
C_{0}(t,t') \equiv \Cov\!\!\left(f(t),f(t')\right) \nonumber \\
C_{1}(t,t') \equiv \Cov\!\!\left(f(t),\dot{f}(t')\right), \quad
\tilde{C}_{1}(t,t') \equiv \Cov\!\!\left(\dot{f}(t),f(t')\right) \\
C_{2}(t,t') \equiv \Cov\!\!\left(\dot{f}(t),\dot{f}(t')\right) \nonumber
\end{eqnarray}
We will also use lowercase $c$ to denote the single point limit of these,
i.e.\ $c_0(t) \equiv C_0(t,t)=\Var\!\left(f(t)\right)$ {\it etc}.

\subsection{Restricting the field to a curve}
\label{sub:restrict}

Before we write down expressions for $C_0$, $C_1$ and $C_2$, let us take note
of special properties of these functions, which follow from the symmetry of 
our field. 
The fields considered are translationally invariant.
This implies that the mean and the
variance of the field are constant over the domain. It allows us to apply
a constant linear transformation $\psi = A\tilde{\psi}+B$, normalizing the
field so that $\langle\psi(\bi{r})\rangle=0$ and 
$c_0 = \langle\psi^2\left(\bi{r}\right)\rangle=1$ 
for all $\bi{r}$ (this is true for 
equations~\eref{eq:randomwave}~and~\eref{eq:bessel}, however we will keep $c_0$
in our expressions below, to make explicit the effect of the multiplicative
scaling).

Furthermore, assuming smoothness of $\psi$, it follows that 
\begin{equation}
 2 \langle\psi(\bi{r}) \triangledown \psi(\bi{r}) \rangle = 
\triangledown\langle\psi^2\left(\bi{r}\right)\rangle = 0. 
\label{eq:psidpsi}
\end{equation}

In general, the restricted fields are not translationally invariant. Therefore
the computation of their correlations requires special attention, and will involve
the geometric properties of the reference curve, such as its curvature. However,
in the limit $t\rightarrow t'$, one can still project~\eref{eq:psidpsi} onto the 
curve, and conclude that
$\tilde{C}_1(t,t') \rightarrow - C_1(t,t')$, and in 
particular $c_1=0$.

The fields we consider here are also isotropic\footnote{An example of
a random field which is \emph{not\/} isotropic, is the non-monochromatic random
wave models used in~\cite{longuethiggins}, which have a finite average 
wavenumber vector (carrier wave)}, so the correlation function has the form
$\langle \psi(\bi{r})\psi(\bi{r}') \rangle = G(|\bi{r}-\bi{r}'|)$.
To make use of this property, we change the coordinates
\begin{equation*}
\bi{u}=\bi{r} + \bi{r}' , \quad \bi{d}=\bi{r} - \bi{r}',
\end{equation*}
so that derivatives with respect to $\bi{u}$ vanish.
Derivatives with respect to
$\bi{d}$ can be expressed in terms of the single parameter function
$G(d)$ (where $d=|\bi{d}|$).

This results in the following expressions for the
correlation functions~\eref{eq:corrdefs}
\begin{eqnarray}
C_{0} = G(d) \nonumber \\
C_{1} = - \dot{\bi{r}'} \cdot \hat{\bi{d}}\, G'(d),\quad 
\tilde{C}_{1} = \dot{\bi{r}} \cdot \hat{\bi{d}}\, G'(d)\\
C_{2} = -\sum_{i,j} \dot{r}_i 
        \left[\left(G'' - \frac{G'}{d}\right) \frac{d_i d_j}{d^2}
              + \frac{G'}{d} \delta_{i,j}
        \right] {\dot{r}'}_j \nonumber
\label{eq:corrcurv}
\end{eqnarray}
where $\bi{r}' \equiv \bi{r}(t')$, and $\hat{\bi{d}} \equiv \bi{d}/d$.

\subsection{The nodal intersections statistics}
\label{sub:inters_stats}

The density of zeros of $f$ takes the form 
\begin{equation}
\rho(t)=\sum_i \delta(t-t_i) = \delta\!\left(f(t)\right)|\dot{f}(t)|,
\quad \textrm{where }f(t_i)=0.
\label{eq:ni_denst}
\end{equation}
 
Following M.~Kac~\cite{kac:prob}, we use the Fourier representation:
\begin{equation*}
\langle \rho(t) \rangle = \frac{1}{2\pi^{2}}
   \int\!\!\!\int_{-\infty}^{\infty}\frac{\rmd\xi \, \rmd\eta}{\eta^{2}}
                         \left\langle \rme^{\rmi\xi f(t)}
                                 ( 1-\rme^{\rmi\eta\dot{f}(t)} ) \right\rangle.
\end{equation*}
This integral can be solved using elementary properties of multinormal 
variables~\eref{eq:multivar_denst}.
For any multinormal variable $\bi{x}$ with zero mean, and real vector $\bi{k}$,
the following identity holds:
\begin{equation}
\fl \langle \rme^{\rmi \bi{k}\cdot\bi{x}} \rangle = 
   \exp\left(-\frac{1}{2} \sum_{i,j} k_i C_{i,j} k_j\right) 
  \quad\textrm{where } C_{i,j}=\Cov(x_i,x_j)
\label{eq:expmulti}
\end{equation}
Using the Gaussianity of the field $f$, this
identity can be applied to the mean density. The resulting integral separates
into simple Gaussian integrals, and we get the following result for the density:
\begin{equation}
\left\langle \rho\right\rangle =\frac{1}{\pi}\sqrt{\frac{c_{2}}{c_{0}}}
\label{eq:density}
\end{equation}

We now move on to two point statistics. When dealing with ``point processes''
 (such as the zeros of $f$),
whose amplitude is a sum of delta functions, $\langle \rho \rho' \rangle$
normally contains a delta function at $t=t'$, corresponding to the strong
autocorrelation. It is customary to subtract this term, to regularize the correlation
function at $t=t'$.  Thus, the two point correlation function is defined 
 (e.g.\ in~\cite{mehta:rand_mat}) as
\begin{eqnarray*}
\fl
R(t,t') &=
   \left\langle \sum_{i\ne j} \delta(t-t_i) \delta(t'-t_j) 
   \right\rangle \nonumber\\
\fl     &=\langle \rho(t)\rho(t') \rangle - \delta(t-t')\langle\rho(t)\rangle .
\end{eqnarray*}

Using the same approach as above, we write:
\begin{equation}
\fl
\langle \rho \rho' \rangle = 
  \left\langle
    \frac{1}{4\pi^{4}}
    \int\frac{\rmd\xi\,\rmd\eta\,\rmd\xi'\,\rmd\eta'}{\eta^{2}\eta'^{2}}
       \rme^{\rmi\xi f}   (1-\rme^{\rmi\eta \dot{f}})
       \rme^{\rmi\xi' f'} (1-\rme^{\rmi\eta' \dot{f}'})
  \right\rangle
\label{eq:corr_fourdef}
\end{equation}

To make use of~\eref{eq:expmulti}, we write down the covariance matrix of the four 
variables $(f,f',\dot{f},\dot{f}')$
appearing in this expression (see definitions in \eref{eq:corrdefs}):

\begin{equation}
M=\left(\begin{array}{cccc}
c_0   & C_{0}         & 0            & C_{1}\\
C_{0} & c_0           & \tilde{C}_{1}& 0    \\
0     & \tilde{C}_{1} & c_{2}        & C_{2}\\
C_{1} & 0             & C_{2}        & c_{2}
\end{array}\right)
\equiv
\left(\begin{array}{cc}
A     & C\\
C^{T} & B\end{array}\right)
\label{eq:amp_corrmat}
\end{equation}
 ($A$,$B$,$C$ stand for the $2\times 2$ submatrices).

Applying \eref{eq:expmulti} to \eref{eq:corr_fourdef} we get more complicated
integrals, which are nevertheless solvable (derivation 
in~\ref{sec:deriv_corr}). The resulting expression for $R(t,t')$ is
\begin{equation*}
R(t,t')=
  \frac{1}{\pi^2}
  \frac{a}{|A|^{3/2}}
  \left(\sqrt{1-\hat{c}^{2}}+\hat{c}\arcsin(\hat{c})\right),
\end{equation*}
where $a=c_2|A|-c_0|C|$, $\hat{c}=(C_2|A|-C_0|C|)/a$, $|A|={c_0}^2-{C_0}^2$
and $|C|=-C_1\tilde{C}_1$ (here $|A|$ and $|C|$ stand for determinants---not
absolute value).

From this, the normalized correlation coefficient 
$\mathcal{R} \equiv R/{\langle \rho \rangle}^2 - 1$, from which we will derive
the other two point statistics, follows immediately:
\begin{equation}
\mathcal{R}(t,t')= 
  \frac{c_0}{c_2} \frac{a}{|A|^{3/2}}
  \left(\sqrt{1-\hat{c}^{2}}+\hat{c}\arcsin(\hat{c})\right) - 1
\label{eq:correl}
\end{equation}

\section{Simple curves in monochromatic random fields}\label{sec:mrw}

The monochromatic Gaussian field~\eref{eq:randomwave}
is often used as a statistical model for eigenfunctions of chaotic
billiards. In this context, the ``semiclassical'' regime (large $k$) has
special significance. In terms of our statistical system, this means that
we would be particularly interested in the $k \gg \kappa$ limit (where
$\kappa$ is the curvature of the reference curve).
In this regime, the first approximation for the curve is a straight line,
and the second one is a circular arc. These two cases are also important
for their simplicity---in both cases, the domain distance $d$ between
two points on the curve depends only on the curve distance $|t-t'|$,
so \eref{eq:corrcurv} assumes a simpler form.

For the straight line, $\dot{\bi{r}}=\hat{\bi{d}}$ is constant and  $d=|t-t'|$.
We get:
\numparts{}
\begin{equation}
\fl
C_0 = G(d) ,\quad C_1 = -G'(d), \quad C_2 = -G''(d)
\label{eq:corrline}
\end{equation}

For a circle of radius $r$, define $\alpha \equiv (t-t')/(2r)$, so
$d=2r\sin\alpha$, and by simple geometrical identities, we get:
\begin{eqnarray}
\fl
C_0 = G(d), \quad C_1 = -\cos\alpha G'(d), \nonumber \\
\fl
C_2 = \sin^2\alpha \frac{G'(d)}{d} - \cos^2\alpha G''(d)
\label{eq:corrcirc}
\end{eqnarray}
\endnumparts{}

\subsection{Normalized correlation for the line and the circle}
\label{sub:rw_corr_circle}
Substituting \eref{eq:corfunrw} in \eref{eq:corrline} and \eref{eq:corrcirc},
we get the amplitude correlations. The results for the circle are
 (the straight line is given by their $\alpha \rightarrow 0$ limit):

\begin{eqnarray}
\label{eq:ampcorr_rw}
C_0 = J_0(kd) \nonumber\\
C_1 = k\cos(\alpha) J_1(kd) \\
C_2 = k^2\left(J_0(kd) \cos^2(\alpha) - \frac{J_1(kd)}{kd}\right) \nonumber
\end{eqnarray}

For the density, we substitute $c_2=C_2(0)=\frac{1}{2}k^2$ in~\eref{eq:density}
getting $\langle\rho\rangle={k}/(\sqrt{2}\pi)$. We use this to select a curve
parameter which measures length in units of average spacing (so we can expect
meaningful results at the $\langle\rho\rangle\rightarrow\infty$
limit): 
\begin{equation*}
s \equiv (t'-t)\langle\rho\rangle = {(t-t')k}/(\sqrt{2}\pi). 
\end{equation*}
In the new units, $\alpha = s\sqrt{2}\pi/(2kr)$ and the argument of the
Bessel functions is
\[ \theta \equiv kd = 2kr\sin\!\!\left(\frac{\sqrt{2}\pi s}{2kr}\right) 
   \xrightarrow[kr\rightarrow\infty]{} \sqrt{2}\pi s
\]

For the NI normalized correlation, we substitute~\eref{eq:ampcorr_rw}
in~\eref{eq:correl}, getting:
\begin{eqnarray}
\fl
\hat{c}=\frac{\left[ J_{0}(\theta)\cos^{2}\alpha-J_{1}(\theta)/\theta \right]
              \left[ 1-{J_{0}(\theta)}^{2} \right]
              - J_{0}(\theta){J_{1}}^{2}(\theta)\cos^{2}\alpha}
             {\left[1-{J_{0}}^{2}(\theta)\right]/2
              -{J_{1}}^{2}(\theta)\cos^{2}\alpha} \nonumber\\
\fl
\mathcal{R} = \frac{1-{J_{0}}^{2}(\theta)-2{J_{1}}^{2}(\theta)\cos^2\alpha}
         { {\left[1-{J_{0}}^{2}(\theta)\right]}^{3/2} }
    \left( \sqrt{1-{\hat{c}}^{2}}+\hat{c}\arcsin\hat{c} \right) -1
\label{eq:rw_covar}
\end{eqnarray}

To get the universal distribution ($k$ independent, high density limit), we 
take $k\rightarrow\infty$ keeping $s$ constant.
\begin{eqnarray}
\fl
\hat{c}=\frac{\left[ J_{0}(\theta)-J_{1}(\theta)/\theta \right]
              \left[ 1-{J_{0}(\theta)}^{2} \right]
              - J_{0}(\theta){J_{1}}^{2}(\theta)}
             {\left[1-{J_{0}}^{2}(\theta)\right]/2
              -{J_1}^2(\theta)} \nonumber\\
\fl
\mathcal{R} = \frac{1-{J_{0}}^{2}(\theta)-2{J_{1}}^{2}(\theta)}
         { {\left[1-{J_{0}}^{2}(\theta)\right]}^{3/2} }
    \left( \sqrt{1-{\hat{c}}^{2}}+\hat{c}\arcsin\hat{c} \right) -1
\label{eq:rw_covar_line}
\end{eqnarray}
Note that in the case of the straight line ($\kappa \equiv 1/r\rightarrow 0$),
we have $\theta = \sqrt{2}\pi s$ and $\alpha=0$ exactly, so there is no $k$
dependence, and the universal function~\eref{eq:rw_covar_line} is exact for any
$k$ (this is expected, because with no boundary and no curvature, the system
has no natural scale).

\begin{figure}[htbp]
\centering
\includegraphics[clip,scale=0.8]{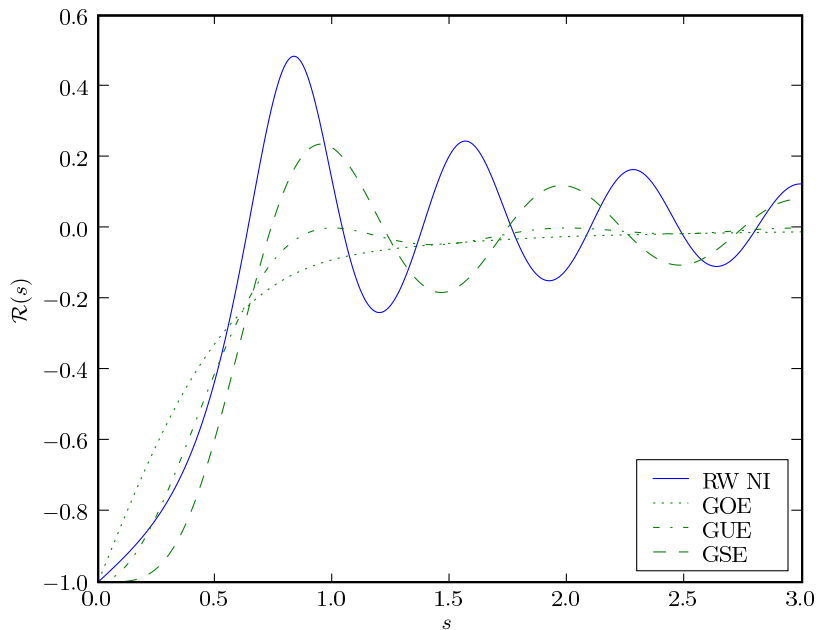}
\caption{\label{cap:covar}Normalized Correlation: NI of RW vs RMT level 
spacings.}
\end{figure}

To further study the universal correlation function~\eref{eq:rw_covar_line}, we
derive its asymptotic expansion. The first few terms in the expansion
of $\mathcal{R}(s)$ near $s=0$ are:
\begin{equation}
-1 + \frac{\pi^{2}}{16}s + \frac{37\pi^{4}}{2304}s^{3}
   + \frac{\pi^{4}}{1296\sqrt{2}}s^{4}.
\label{eq:corr_small}
\end{equation}
This rises linearly from $-1$, with a smaller slope than the GOE ensemble (other
RMT ensembles are not linear at this limit), however, as is shown
in~\fref{cap:covar}, it rises quicker to positive values, resulting in a smaller
range of high repulsion.

For large values of $s$, we find decaying terms, oscillating in frequencies
which are multiples of the dominant frequency 
$\omega \equiv 2\pi \sqrt{2}$. The asymptotic expansion has the form:
\begin{equation}
\fl \mathcal{R}(s)\sim \Re \sum_{n=1}^{3} \frac{1}{{(\pi\omega s)}^{n}}
            \sum_{m=0}^n
               q_{n,m}(\rmi\pi)
               \rme^{\rmi m(\omega s-\pi/2)}
              + \Or(s^{-4})
\label{eq:corr_big}
\end{equation}
where $q_{n,m}(x)$ are the following polynomials:
\begin{eqnarray*}
\fl q_{1,0}=1, \quad q_{1,1}=9 \\
\fl q_{2,0}=\frac{25}{4}, \quad q_{2,1}=\frac{49}{3}+\frac{39}{2}x, \quad
            q_{2,2}=\frac{121}{12}\\
\fl q_{3,0}=\frac{169}{4}+\frac{17}{2}x^2, \quad
            q_{3,1}=\frac{1369}{24}+\frac{511}{6}x+\frac{205}{8}x^2, \\
\fl \qquad  q_{3,2}=\frac{1681}{60}+\frac{55}{12}x, \quad
            q_{3,3}=\frac{529}{40}
\end{eqnarray*}

Writing the leading terms explicitly, we have
\begin{equation}
\mathcal{R} \sim \frac{1}{\pi\omega s}\left(1+9\sin(\omega s)\right)
\label{eq:rw_correl_lead}
\end{equation}

This $s^{-1}$ decay is slower than the decay of random matrix ensembles
 ($s^{-2}$ for GOE and GUE, and $s^{-1} \times [\mathrm{oscillating\ part}]$  
for GSE). The slow decay has a considerable effect on all statistics
that probe large distances, as will be shown in~\sref{sub:var_power}.

\begin{figure}[htbp]
\centering
\includegraphics[clip,scale=0.7]{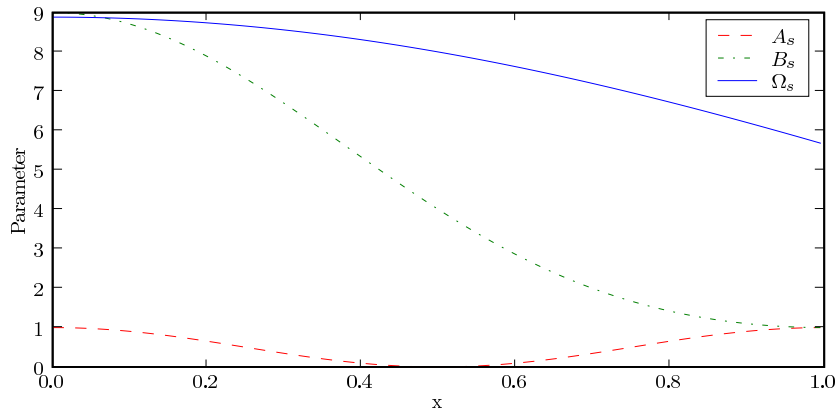}
\caption{\label{cap:circ_rwcorr_coefs}
Asymptotic coefficients and frequency for RW NI on half a circle as a function
of $x=\sqrt{2}s/(kr)$.}
\end{figure}

When we look at intersections on a full circle, $s$ goes over all values, from
$-kr/\sqrt{2}$ to $kr/\sqrt{2}$. This range contains values large enough to
make \eref{eq:rw_covar_line} unsuitable for approximating the correlations. The
leading terms in the asymptotic expansion of \eref{eq:rw_covar}  
 (the equivalent of \eref{eq:rw_correl_lead}) are found to be:

\begin{equation}
\mathcal{R}(s) = 
  \frac{1}{\pi\Omega_{s}s}\left(A_{s} + B_{s}\sin(\Omega_{s}s)\right) 
  + \Or(s^{-2})
\label{eq:rw_correl_lead_circ}
\end{equation}
where 
\begin{eqnarray*}
A_{s}=\cos^{2}\!\!\left(\frac{2\pi s}{\sqrt{2}kr}\right),
B_{s}={\left[2+\cos\!\!\left(\frac{2\pi s}{\sqrt{2}kr}\right)\right]}^{2}, \\
\Omega_{s} = \sinc\!\!\left(\frac{s}{\sqrt{2}kr}\right) \cdot \omega
\end{eqnarray*}
 (with normalized $\sinc(x)\equiv \sin(\pi x)/(\pi x)$), are slowly varying
functions, starting from $A_0=1$,$B_0=9$ and $\Omega_0=\omega$ (as expected
from~\eref{eq:rw_correl_lead}, which holds for $s\ll kr$), and ending
in $A_{\pm S}=1$,$B_{\pm S}=1$ and $\Omega_{\pm S}=4\sqrt{2}$ for 
$S=kr/\sqrt{2}$. The functions $A_s$, $B_s$ and $\Omega_s$
are illustrated in~\fref{cap:circ_rwcorr_coefs}.
As we increase $k$, the rate of change becomes slower, but the
normalized length of the curve increases, and the value of the coefficients for
maximally separated points ($|s|=kr/\sqrt{2}$) remains the same.

Before proceeding to compute other statistics, we would like to point out that
the oscillations in $p(s)$ (see~\sref{sub:nn_dist}) have a frequency which is
numerically consistent with half the dominant frequency 
$\omega \equiv 2\pi \sqrt{2}$.

\subsection{Number variance and power spectrum}
\label{sub:var_power}

The variance of the number of intersections in a finite segment of
length $L$ is often used as a measure for the ``rigidity'' of the distribution
 (a completely rigid set of points will have zero variance).
The power spectrum~\cite{rice:rnoise1}
measures periodicity in the distribution (which is also manifested by 
oscillations in the correlation function). These statistics are given by
integral transforms of $\mathcal{R}$ which are hard to solve directly. Instead,
the asymptotic expansions of the correlation function can be used to expand them
in series at regions of interest, as explained in~\ref{sec:asymp_integs}.

The variance for the number of intersections on a finite segment of length $L$
can be calculated from the normalized correlation by~\cite{mehta:rand_mat}:
\begin{equation}
\Sigma^2(L)=L+2\int_{0}^{L}(L-s)\mathcal{R}(s)\rmd s
\label{eq:numvar}
\end{equation}
We can use~\eref{eq:corr_small} and~\eref{eq:corr_big} to find the asymptotic
behaviour of this statistics.
For small $L$, we find
\[
\fl \Sigma^2 \sim L-L^{2}+\frac{\pi^{2}}{48}L^{3}
           +\frac{37\pi^{4}}{23040}L^{5}
           +\frac{\pi^{4}}{19440\sqrt{2}}L^{6}
\]
The first two terms follow from the formal definition of $\Sigma^2$, and therefore
coincide with the corresponding terms in the RMT ensembles. The third
term corresponds to the probability of finding two intersections in the small
interval $[0,L]$. It is of order $L^3$, as in GOE (again with a smaller 
factor).

For large $L$, we find
\begin{eqnarray}
\fl \Sigma^2 \sim& \frac{2q_{1,0}}{(\pi\omega)}(L\ln L-L)
      + L(1+2\mathcal{M}^{0}_{\mathrm{RW}}) \nonumber\\
\fl & -\frac{2q_{2,0}}{{(\pi\omega)}^{2}}(\ln L+1)
      - 2\mathcal{M}^1_{\mathrm{RW}} \nonumber\\
\fl & + \frac{1}{{(\pi\omega)}^3L} 
        \Re\!\left[q_{3,0}(\rmi\pi)
            +2\rmi\pi^2 q_{1,1}\rme^{\rmi\omega L}\right]
      + \Or(L^{-2})
\label{eq:sigma2_big}
\end{eqnarray}
where $\mathcal{M}^0_{\mathrm{RW}}$ and $\mathcal{M}^1_{\mathrm{RW}}$ 
are fixed constants, given by
\begin{eqnarray*}
\fl \mathcal{M}^0_\mathrm{RW} = &\int_0^1 \mathcal{R}(s)\rmd s 
        +\int_1^\infty\left[ \mathcal{R}(s)-\frac{1}{\pi\omega s}\right] \rmd s 
      \sim -0.336 \\
\fl \mathcal{M}^1_\mathrm{RW} = 
        &\frac{1}{\pi\omega}\left(9\frac{\cos \omega}{\omega} -1\right)
        +\int_0^1 s \mathcal{R}(s)\rmd s \\
\fl     &+\int_1^\infty\left[ s \mathcal{R}(s)-\frac{1}{\pi\omega}(1+9\sin\omega s)
                            -\frac{25}{4{(\pi\omega)}^2 s} \right] \rmd s \\
\fl     &\sim -0.0826
\end{eqnarray*}

\begin{figure}[htbp]
\centering
\includegraphics[clip,scale=0.8]{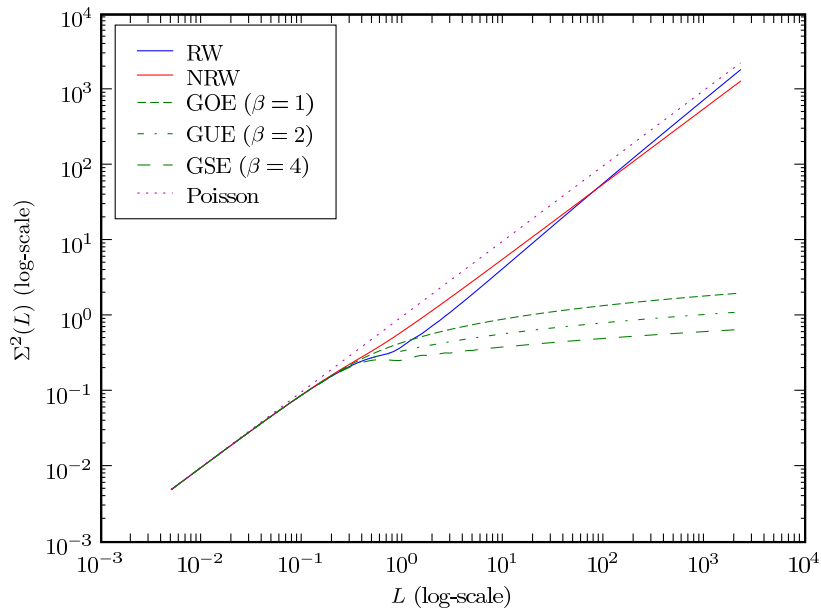}
\caption{\label{cap:numvar}Number variance in segments of length $L$. 
For large $L$, the NRW curve is linear, hence parallel to the Poisson curve.
The RW curve rises faster (the crossing point is at $L\sim 32\times 10^3$,
beyond the range of this plot).}
\end{figure}

Writing down the first terms of~\eref{eq:sigma2_big} explicitly, we have:
\begin{eqnarray}
\fl \Sigma^2(L)\sim& \frac{2}{\pi\omega}L\ln L 
                    +\left(1-\frac{2}{\pi\omega}
                                      +2\mathcal{M}^0_{\mathrm{RW}}
                      \right) L  \nonumber\\
\fl                & -\frac{25}{{(2\pi\omega)}^2}\ln L
                    -2\left(\mathcal{M}^1_{\mathrm{RW}}
                            +\frac{25}{{(2\pi\omega)}^2}\right)
\end{eqnarray}
Note that this variance is asymptotically larger than that of the random 
 (Poisson) distribution (where $\Sigma^2 \sim L$).
However, due to the numerical constants, the $L\ln L$ term dominates only when
$L> \exp(\pi\omega(\mathcal{M}^0_{\mathrm{RW}}+1/2)-1) \sim 36$, 
and only at $L\sim 32\times 10^3$ does it become larger than the Poisson 
variance.
At segments whose length is a few average spacings, the variance already
becomes significantly larger than the ``rigid'' RMT ensembles.
In \fref{cap:numvar}, the variances of several distributions are compared.

It should be noted that in the case of independent site percolation (for grids
with bounded, convex cells), a straight line passes only once through each cell.
Hence the sign changes along the line form a Poisson process, with linear number
variance. The fact that the asymptotic behaviour of the random wave model 
behaves differently is of interest, because the predictions of the percolation
model~\cite{bogomolny:percolation} matched the numerical data for other 
parameters which characterize the behaviour of RW nodal lines in the high energy
limit. Namely, the nodal count (number of nodal domains contained inside a given
area)~\cite{blum:nd_stats} and the SLE driving force~\cite{bogomolny:sle}.

\begin{figure}[htbp]
\centering
\includegraphics[clip,scale=0.8]{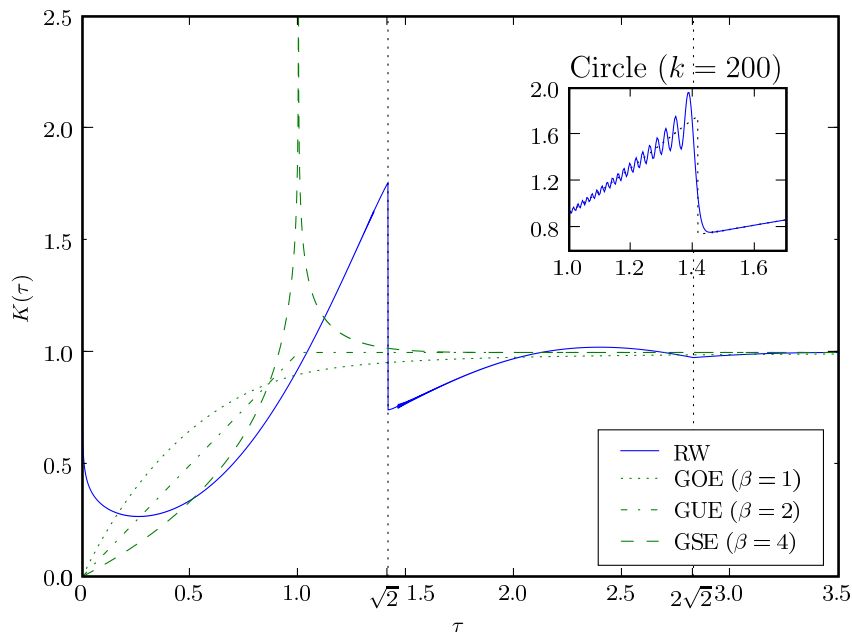}
\caption{\label{cap:rw_formfact}Form factor for RW Nodal Intersections.
The logarithmic divergence to $+\infty$ at $\tau=0$ is too steep to be observable
in this scale.}
\end{figure}

The form factor (scaled power spectrum) for the intersections is the Fourier 
transform of the scaled correlation function (up to the subtraction of a 
$\delta(\tau)$ term),
\begin{eqnarray}
K(\tau)&=\int_\Gamma \rme^{\rmi 2\pi\tau s} 
      \frac{\langle \rho(x)\rho(x+s)\rangle}{{\langle \rho \rangle}^2}
      \rmd s - \delta(\tau) \nonumber\\
       &= 1+ \int_\Gamma \rme^{\rmi 2\pi\tau s}\mathcal{R}(s) \rmd s.
\label{eq:formfact}
\end{eqnarray}
In~\fref{cap:rw_formfact} this is plotted and compared to the corresponding
RMT values.

By~\eref{eq:rw_correl_lead} it is clear that $K$ diverges for $\tau=0$
 (since $1/(\pi\omega s)$ is non integrable). Similarly, from the form of
equation~\eref{eq:corr_big}, we expect singularities at integer multiples of
$\sqrt{2}$ (corresponding to the angular frequency $\omega=2\pi\sqrt{2}$). 
To quantify this, we use the asymptotic expansion of $\mathcal{R}$ to expand
$K(\tau)$ in small regions around the singular points $\tau=\sqrt{2}n+\delta$
for $\delta\ll 1$.

For the divergence at 0 $(\tau \ll 1)$, we get
\[
\fl K(\tau) \sim 1+2\mathcal{M}^0_{\mathrm{RW}}
             -\frac{2[\gamma+\ln(2\pi)]}{\pi\omega}
             -\frac{2}{\pi\omega}\ln\tau
\]
This is a logarithmic divergence. It drops down to 1 at 
$\tau_0 \sim 8\times 10^{-6}$ (which is too small to be observable 
in \fref{cap:rw_formfact}). The parameter $K(0)=1+\int_\Gamma \mathcal{R}(s)\rmd s$
can be viewed as a measure of asymptotic spectral rigidity~\cite{bohigas:rmt}.
In our case, we see that we have ``infinite softness''---the slow decay of the
correlations outweighs the short distance repulsion.

One notable feature of \fref{cap:rw_formfact}, is the finite jump at 
$\tau=\sqrt{2}$.
Expanding for this region, we find
\begin{eqnarray*}
\fl K\!\!\left(\sqrt{2}+\delta\right) \sim & 
     A + B\delta \\
\fl &-\frac{9}{2\omega}\sign[\delta] - \frac{39}{16\pi}|\delta|
     +\frac{49}{12\pi^3}\delta\ln|\delta|
\end{eqnarray*}
where $A\sim1.25$ and $B\sim1.53$ are numerical constants.

For the case where the reference line is a circle (inset 
of~\fref{cap:rw_formfact}), we observe strong oscillations in the form factor
for $\tau<\sqrt{2}$.
These oscillations are caused by the ``drifting'' of the frequency $\Omega_s$
in~\eref{eq:rw_correl_lead_circ}.
Their frequency increases for large values of $k$, and we would expect them
to cancel out when averaged over a wide enough window of frequencies.

\section{The normal derivative of a Gaussian field}\label{sec:nderiv}

When dealing with solutions of boundary value problems (e.g.\ two dimensional
quantum billiards), the boundary of the domain is a curve that plays a special
role. It is therefore worthwhile to investigate its intersections with the
nodal lines. For a Dirichlet boundary condition, the boundary is part of the 
nodal set, and the derivative of the eigenfunction in the tangent direction is 
zero. At the points where another nodal line intersects the boundary, the 
derivative in the direction of the other line will also vanish.
Since nodal lines of solutions
of the Helmholtz equation intersect at right angles, the intersections we wish
to explore are also zeros of the \emph{normal derivative\/} of the field with
respect to the curve. This fact makes the normal derivative of other fields,
such as the unbound random wave, natural candidates for comparison with the
statistics of such intersections.

If fact, the normal derivative of a Gaussian random wave field is directly 
related to the ``boundary modified'' field mentioned in~\sref{sec:intro}.
The boundary modified random waves~\cite{berry:nodalstat}, with correlation
function approaching~\eref{eq:corfunndrw}, might be realized by symmetrization
of the unbound random wave field~\eref{eq:randomwave}:
$\psi_\mathrm{BRW}(x,y) \equiv \psi(x,y)-\psi(x,-y)$ 
 (where $\psi=\psi_\mathrm{RW}$ of~\eref{eq:corfunrw}). In this case,
the correlations on the boundary itself can be defined as the limit of 
correlations on a line parallel to the $x$ axis at $y\rightarrow 0+$. To
get a meaningful limit we must introduce a $y$-dependent scaling, to keep 
$\langle {\psi_\mathrm{BRW}}^2 \rangle$ constant as we approach $y=0$.
Denoting $\phi_y(x)\equiv \psi_\mathrm{BRW}(x,y)$, the average square
amplitude for a line at height $y$ is:
\[
\langle {\phi_y(x)}^2 \rangle =
  \langle {\left( \psi(x,y) - \psi(x,-y) \right)}^2 \rangle
  = 2\left(c_0 - C_0(2y)\right)
\]
The correlation function $C_0$ is symmetric, so in the generic case (and,
in our case specifically), its value for small $d$ can be approximated by
a square function $C_0(d)\sim c_0 + \frac{1}{2}{c''}_0 d^2$.
Thus, $\langle {\phi_y(x)}^2 \rangle \sim {(2y)}^2 |{c''}_0|$ (note that
${c''}_0$ is negative), and the scaled field will be:
\begin{eqnarray*}
\fl \widehat{\phi_y}(x) = 
   \frac{\psi(x,y)-\psi(x,-y)}
        {\sqrt{\langle {\phi_y(x)}^2\rangle}} \nonumber \\
\fl \sim
   \frac{1}{\sqrt{|{c''}_0|}} \frac{\psi(x,y)-\psi(x,-y)}{2y}
   \xrightarrow[y\rightarrow 0+]{}
   \frac{1}{\sqrt{|{c''}_0|}} \frac{\partial\psi(x,y)}{\partial y} .
\end{eqnarray*}
Therefore, the properly scaled boundary modified field is
proportional to the normal derivative of the random wave field used to generate
it.

\subsection{Amplitude correlations of the normal derivative on the reference curve}
\label{sub:bm_restrict}

The normal derivative of the field across the curve is given by
$g(t)=\bi{n}(t)\cdot\bi{\triangledown}\psi(\bi{r}(t))$, where 
$\bi{n}(t)$
is the unit normal vector of the curve at point $t$. Repeating the computation
described in~\sref{sub:restrict}, we can
calculate its correlation functions.

\begin{eqnarray}
\fl \left\langle g(t) g(t')\right\rangle &=
      \sum_{i,j} n_{i} 
         \frac{\partial^{2} 
                  \left\langle \psi\!\!\left(\bi{r}(t)\right)
                       \psi\!\!\left(\bi{r}(t')\right) \right\rangle}
              {\partial r_i \partial {r'}_{j}} {n'}_{j} \nonumber \\
\fl &= 
   \sum_{i,j}
    -n_{i}
    \left[ \left(G''(d)-\frac{G'(d)}{d}\right) \frac{d_{i}d_{j}}{d^{2}} 
           + \frac{G'(d)}{d} \delta_{ij} \right]
    {n'}_{j}
\label{eq:curv_ndrv_c0}
\end{eqnarray}

For the circle (as before $\alpha=(t-t')/(2r)$, $d=2r\sin \alpha$), we have the
identities $\bi{n}\cdot\bi{n}'=\cos(2\alpha)$ and
$\bi{n}'\cdot\hat{\bi{d}} = -\bi{n}\cdot\hat{\bi{d}} = \sin\alpha$.
Inserting these in~\eref{eq:curv_ndrv_c0} we get:

\numparts{}
\begin{equation}
\fl C_0 = \left\langle g g' \right\rangle =
    -\cos^2\alpha \frac{G'(d)}{d} + \sin^2\alpha G''(d)
\label{eq:circ_ndrv_c0}
\end{equation}

The other two correlation functions are calculated in a similar manner (for a
general curve) in~\ref{sec:curv_ndrv_ampcorr}. For the case where the curve is
a circle, they assume the following form\footnote{This form was chosen to make
the $\alpha\rightarrow 0$ limit more evident. The forms 
in~\ref{sec:curv_ndrv_ampcorr} are more suitable for calculations involving
Bessel functions or generic derivatives.}:

\begin{eqnarray}
\fl C_1 = \langle g\dot{g}'\rangle = &
    \cos\alpha \left( 1-3\sin^2\alpha \right)
       \left( \frac{G''}{d} - \frac{G'}{d^{2}} \right) \nonumber\\
\fl  & -\sin^{2}\alpha\cos\alpha
       \left( G^{(3)} + 4\frac{G'}{d^{2}} \right) \\
\fl C_2 = \langle \dot{g} \dot{g}'\rangle = &
      -2\left( \cos(2\alpha) + \frac{7}{8}\sin^{2}(2\alpha)\right)
          \frac{1}{d^{2}} \left( G''-\frac{G'}{d}\right) \nonumber\\
\fl & + \left( 1 - \frac{3}{2}\sin^{2}(2\alpha) \right) 
          \frac{G^{(3)}}{d} \nonumber\\
\fl & -\sin^{2}\alpha \left( \cos^{2}\alpha G^{(4)}
                            + 4\cos(2\alpha) \frac{G'}{d^{3}} \right)
\label{eq:circ_ndrv_c2}
\end{eqnarray}
\endnumparts{}

The correlations for the straight line are easily derived from
equations~\eref{eq:circ_ndrv_c0}--\eref{eq:circ_ndrv_c2} by taking the
limit $\alpha\rightarrow 0$:

\numparts{}
\begin{eqnarray}
 C_0 = - \frac{G'}{d} \label{eq:line_ndrv_c0}\\
 C_1 = \frac{G''}{d} - \frac{G'}{d^{2}} \\
 C_2 = \frac{G^{(3)}}{d} 
       - \frac{2}{d^{2}} \left( G'' -\frac{G'}{d} \right)
\label{eq:line_ndrv_c2}
\end{eqnarray}
\endnumparts{}

To study the normally derived field corresponding to the monochromatic 
random ensemble of~\sref{sec:mrw}, we apply the results above to the
specific correlation function, and discuss the resulting statistics. For 
simplicity, we discuss the case of a straight line first, and follow with
comments about the main differences in the case of a circular curve.

For the straight line, the amplitude correlations are derived from
\eref{eq:line_ndrv_c0}--\eref{eq:line_ndrv_c2}. To get a standard scaling with 
$c_0=1$, we first multiply the field by $\sqrt{2}/k$, getting $G=2J_0(kd)/k^2$, 
and the following correlations for the normal derivative:

\begin{eqnarray}
 C_0 = \frac{2J_1(kd)}{kd} \nonumber \\
 C_1 = k \frac{2J_2(kd)}{kd} \nonumber \\
 C_2 = k^2 \frac{2}{kd} \left( \frac{J_2(kd)}{kd}-J_3(kd) \right)
\label{eq:line_ndrv_rw}
\end{eqnarray}

\subsection{Nodal intersection statistics for a straight line}

We proceed as in \sref{sec:mrw}. First, the density of intersection is
calculated using~\eref{eq:density}:
\begin{equation*}
\langle\rho\rangle = \frac{1}{\pi}\sqrt{\frac{c_{2}}{c_{0}}} =
     \frac{k}{2\pi}
\end{equation*}
We use this to rescale the distance to units of average spacing 
$s=(t-t')k/(2\pi)$, so the argument of the Bessel functions 
in~\eref{eq:line_ndrv_rw} becomes $\theta = 2\pi s$. Substituting these results
in~\eref{eq:correl}, we get:

\begin{eqnarray}
\fl \hat{c} =
  \frac{8}{\theta^2} \frac{(J_2-\theta J_{3})(\theta^2-4{J_1}^2)
                           -4\theta J_1{J_2}^2}
                          {\theta^2-4{J_1}^2-16{J_2}^2} \nonumber \\
\fl \mathcal{R} = 
   \frac{\theta (\theta^{2}-4{J_1}^2-16{J_2}^2)}
        {{(\theta^2-4{J_1}^2)}^{3/2}}
   \left(\sqrt{1-\hat{c}^{2}}-\hat{c}\arcsin\hat{c}\right)-1
\label{eq:ndrv_covar_line}
\end{eqnarray}

\begin{figure}[htbp]
\centering
\includegraphics[clip,scale=0.8]{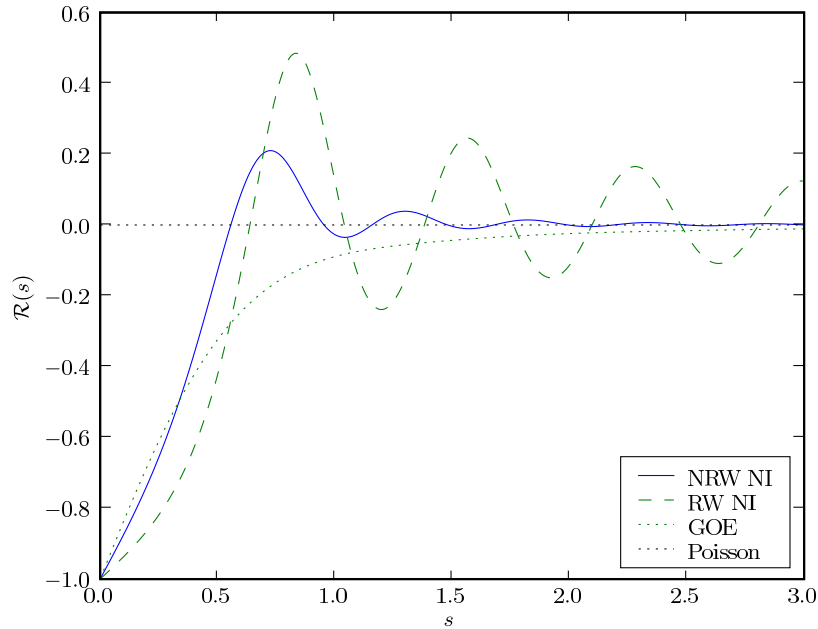}
\caption{\label{cap:corrs_nrw}Normalized Correlation: NI of normally derived
 RW.}
\end{figure}

The asymptotic expansion of $\mathcal{R}$ for small $s$ is:
\[
\mathcal{R}(s) \sim -1 + \frac{\pi^2}{8}s+\frac{13\pi^4}{576}s^3 
          +\frac{\pi^4}{324}s^4
\]
As before, this grows linearly from $-1$. The slope is larger than that of the
unmodified RW, but smaller than the slope of the GOE correlations. As shown in
\fref{cap:corrs_nrw}, it rises quickly to positive values.

The expansion for large $s$ is:
\begin{eqnarray}
\fl \mathcal{R}(s) \sim 
        & \frac{2}{\pi{(2\pi s)}^{3}} \left( 9-25\sin(4\pi s) \right)
         + \frac{-555}{2\pi {(2\pi s)}^4} 
               \cos(4\pi s) \nonumber\\
\fl     &+ \frac{1}{2\pi {(2\pi s)}^5}
               \left(-\frac{69}{2}+\frac{12789}{8}\sin(4\pi s)\right).
\label{eq:ndrv_correl_big}
\end{eqnarray}

$\mathcal{R}(s)$ decays as $s^{-3}$, which is faster than the decay of the
correlations for the corresponding RMT ensembles. As
in~\eref{eq:corr_big}, each asymptotic term has oscillating parts, with
frequencies which are multiples of the dominant frequency
$\omega_\mathrm{NRW}=4\pi$. As in~\sref{sub:rw_corr_circle}, we find that
the oscillations in the nearest neighbour density of the NRW, described in
\sref{sub:nn_dist}, have a frequency which is numerically consistent with half
the dominant frequency $\omega \equiv 4\pi$.

With the $s^{-3}$ decay of the correlations, both $\int \mathcal{R}(s)\rmd s$ and 
$\int s \mathcal{R}(s)\rmd s$ are finite,
so from~\eref{eq:numvar} it is evident that $\Sigma^2$ should increase linearly
for large values of $L$. The asymptotic expansion calculated 
from~\eref{eq:ndrv_correl_big} is:
\begin{equation}
\Sigma^2 \sim (1 + 2\mathcal{M}^0_\mathrm{NRW}) L 
              -2\mathcal{M}^1_\mathrm{NRW}
\label{eq:ndrv_numvar_big}
\end{equation}
with
\[
\mathcal{M}^0_\mathrm{NRW} = \int_0^\infty \mathcal{R}(s)\rmd s \sim -0.2582
\]
and
\[
\mathcal{M}^1_\mathrm{NRW} = \int_0^\infty s \mathcal{R}(s)\rmd s \sim -0.00617 .
\]
As seen in~\fref{cap:numvar}, the asymptotic variance is larger than RMT
ensembles, but smaller than Poisson.

Much in the same way, from~\eref{eq:formfact}, it is evident that the form 
factor
will now be finite at $\tau=0$, and will be continuously differentiable for 
$\tau>0$. Irregularities appear in second and higher derivatives at multiples
of the dominant frequency $\omega_{\mathrm{NRW}}$ 
 (i.e.\ $\tau_n=n \omega_{\mathrm{NRW}}/(2\pi) = 2n$).
The asymptotic expansion for $\tau \ll 1$ is given by:
\begin{eqnarray*}
\fl K(\tau) \sim & 
      (1+2\mathcal{M}^0_\mathrm{NRW})
      +\frac{9}{\pi^{2}}\tau^{2}\ln\tau  \\
\fl & +\tau^2\left[\frac{9}{\pi^2}(\gamma-\frac{3}{2}+\ln 2\pi) 
                   -{(2\pi)}^2\mathcal{M}^2_\mathrm{NRW} \right]
\end{eqnarray*}
with
\[
\fl \mathcal{M}^2_\mathrm{NRW} = 
    \int_0^1 s^2 \mathcal{R}(s)\rmd s 
    + \int_1^\infty s^2\left(\mathcal{R}(s) -\frac{9}{4\pi^4 s^3}\right) \rmd s
  \sim 0.00579
\]

We now have a finite $K(0)\sim 0.4836$. This value is smaller than
the $K=1$ of the Poisson distribution, but larger (softer) than the RMT
ensembles, for which $K(0)=0$.

\begin{figure}[htbp]
\centering
\includegraphics[clip,scale=0.8]{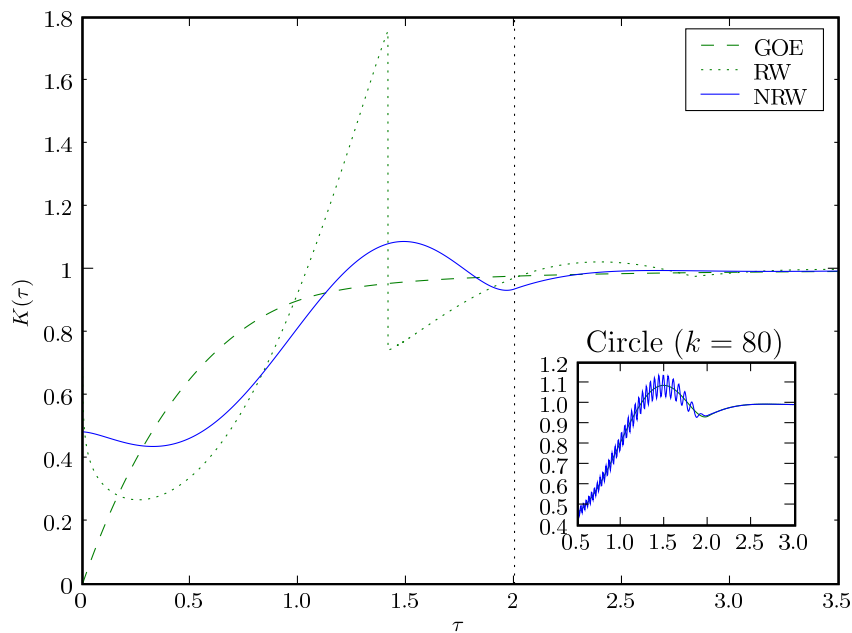}
\caption{\label{cap:formfact_nrw}
Form factor for the NI of normally derived RW.}
\end{figure}

\subsection{Statistics on a circular arc}
\label{sub:nrw_circ}

The case where the reference curve is a circle can be treated similarly. The
amplitude correlations of the monochromatic waves (corrected form 
for~\eref{eq:line_ndrv_rw}) can be calculated directly 
from~\eref{eq:ndrv_curv_sym}.
It should be noted that the amplitude variance $c_0=C_0(0)$ remains 
$\frac{1}{2}k^2$ (as for the straight line). However, for any finite $k$, the
variance of the derivative ($c_2$) is different than the value obtained for the
straight line, and induces a larger density of intersections.

The corrected density of intersections for the circle is:
\[
\fl \rho \equiv \frac{\tilde{k}}{2\pi} = 
   \frac{\sqrt{k^2+{(2\kappa)}^2}}{2\pi}
   \sim \frac{k}{2\pi}\left[1+2{\left(\frac{\kappa}{k}\right)}^2\right]
\]
The $\Or(\kappa^2)$ increase in density, which is due to curvature,
affects the choice of
the scaled curve parameter $s=\rho\cdot(t-t')$ and thus all the correlation
formulae. It is convenient to express the resulting statistics in terms of the
dimensionless parameter
\[ \beta = \frac{k}{\tilde{k}} = \frac{kr}{\sqrt{{(kr)}^2+4}}, \]
which goes to 1 when $kr\rightarrow\infty$. With this, the small $s$
expansion of the normalized correlation is:
\begin{eqnarray*}
\fl \mathcal{R} \sim 
     & -1 + \frac{\pi^2}{8}\beta^2(11-10\beta^2)s \\
 \fl & +\frac{\pi^4}{576}\beta^2(-24+357\beta^2-820\beta^4+500\beta^6)s^3
\end{eqnarray*}

For large values of $s$ (the circle is finite, and $|s|$ can only get values 
up to $\frac{1}{2}\tilde{k}r={(1-\beta^2)}^{-1/2}$), the constant dominant
frequency $\omega=4\pi$ of~\eref{eq:ndrv_numvar_big} is replaced with a 
drifting frequency (changing very slowly with $s$) 
\[ \Omega_s = 4\pi \beta\sinc\!\!\left(\frac{s}{\tilde{k}r}\right), \]
similar to the situation in~\sref{sub:rw_corr_circle}. However, in the case of
the normally derived field, there is another notable difference between the
circular and the straight curve. Namely, for any finite value of 
$kr$, we get $\Or(s^{-1})$ and $\Or(s^{-2})$ terms, which do not appear in the
case of a straight line. The leading term of the normalized correlation
for $s\gg 1$ is
\begin{equation*}
\fl 4\sin^4\alpha[{(4\beta^2\cos^2\alpha-1)}^2 
                  -{(4\beta^2\cos^2\alpha+1)}^2 \sin(\Omega_s s)]\cdot
    \frac{1}{\pi\Omega_s s},
\end{equation*}
where $\alpha=\pi s/(\tilde{k}r)$ is slowly varying up to $\pi/2$. When we are
in the $1\ll s \ll \tilde{k}r$ region, $\alpha$ is very small, and this term
is $\Or(\alpha^4)$. Hence, it vanishes in the straight line limit. The
$\Or(s^{-2})$ term, contains the frequencies $0$, $\Omega_s$ and 
$2\Omega_s$. In the low $\alpha$ region, it is 
$\Or\!\!\left( \alpha^2\cos(\Omega_s s){(\pi\Omega_s s)}^{-2} \right)$ (again, 
vanishes for the straight line). The $\Or(s^{-3})$ term is the first one that
does not vanish when $\alpha\rightarrow 0$.

\begin{figure}[htbp]
\centering
\includegraphics[clip,scale=0.8]{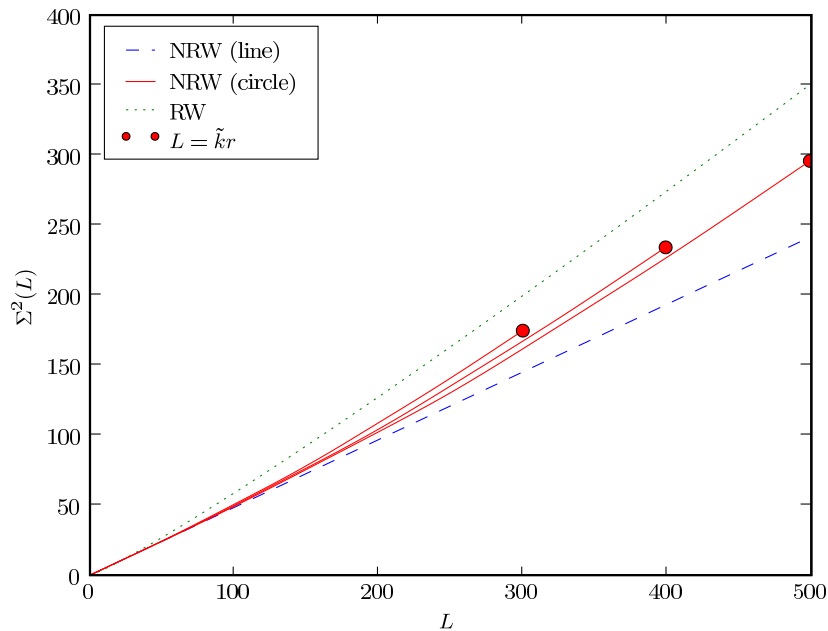}
\caption{Number variance for zeros of NRW on a circular curve, for three
different values of $kr$. The length of the segment is bound by the total
length of the circle, marked by circular dots on the plot.}
\label{cap:var_nrw_circ}
\end{figure}

The effect of these terms is manifested by the number variance statistic.
In~\fref{cap:var_nrw_circ}, the number variance is plotted for three values of
$kr$. As long as $L\ll\tilde{k}r$, the variance stays close to the values 
corresponding to a straight curve given by~\eref{eq:ndrv_numvar_big}.
However, when the angle covered by the segment becomes non negligible,
$\Or(\alpha^4s^{-1})$ and $\Or(\alpha^2s^{-2})$ terms of the correlation become
important, and the variance increases above the linear asymptote.

\section{Comparison with other fields}\label{sec:other}

The statistics considered above can be evaluated numerically for the nodal
intersections of any scalar field with a given reference curve. If a random
Gaussian model is given for the field, the techniques demonstrated above can
be used to extract the theoretical functions predicted by the model, and those
could be compared to the numerical results.
This gives a useful tool for verifying statistical models for such systems.
For the case of chaotic billiards, the numerical results can be compared to the
predictions of random wave models. Also, we can use these statistics to compare
other random wave models with the monochromatic random waves. 
In~\cite{foltin:linedist}, the field with correlation~\eref{eq:corfunsrf} was
compared to the random waves model with respect to special statistics proposed
there (which were related to the probability of finding a nodal line inside a
reference tube). The nodal intersection statistics provide another way to
compare these models.

\subsection{The Gaussian short range field}

The short range field~\eref{eq:corfunsrf} was chosen so that its short distance
behaviour will be the same as that of the monochromatic random wave field, so we
expect the average level spacing to be $\sqrt{2}\pi/k$. Substituting the
correlation function~\eref{eq:corfunsrf} in~\eref{eq:corrcirc}, and denoting
$G=\exp(-\theta^{2}/4)$, $\theta=kd$, we get:
\begin{eqnarray*}
\fl C_0 = G,\quad C_1= \frac{1}{2}k\theta \cos(\alpha) G, \\
\fl C_2 = \frac{1}{2}k^2[\cos(2\alpha)-\frac{1}{2}\theta^2\cos^2(\alpha)]G
\end{eqnarray*}
 (so $c_2=k^2/2$ as expected, and $\theta\rightarrow\sqrt{2}\pi s$ as before).
The nodal intersections correlation for the straight line becomes:
\begin{eqnarray}
\fl \hat{c}=
  \rme^{-\frac{1}{4}\theta^{2}}
        \frac{1-\rme^{-\frac{1}{2}\theta^{2}}-\frac{1}{2}\theta^{2}}
             {1-\rme^{-\frac{1}{2}\theta^{2}}
              -\frac{1}{2}\theta^{2}\rme^{-\frac{1}{2}\theta^{2}}} \nonumber\\
\fl \mathcal{R}(s) =
  \frac{1-\rme^{-\frac{1}{2}\theta^{2}} 
        - \frac{1}{2}\theta^{2} \rme^{-\frac{1}{2}\theta^{2}}}
       {{(1-\rme^{-\frac{1}{2}\theta^{2}})}^{\frac{3}{2}}}
  \left(\sqrt{1-\hat{c}^{2}} + \hat{c}\arcsin(\hat{c})\right)-1
\label{eq:short_covar_line}
\end{eqnarray}

\begin{figure}[htbp]
  \centering
  \subfloat[Correlations]{\label{cap:srf_correl:corrs}
    \includegraphics[clip,scale=0.5]{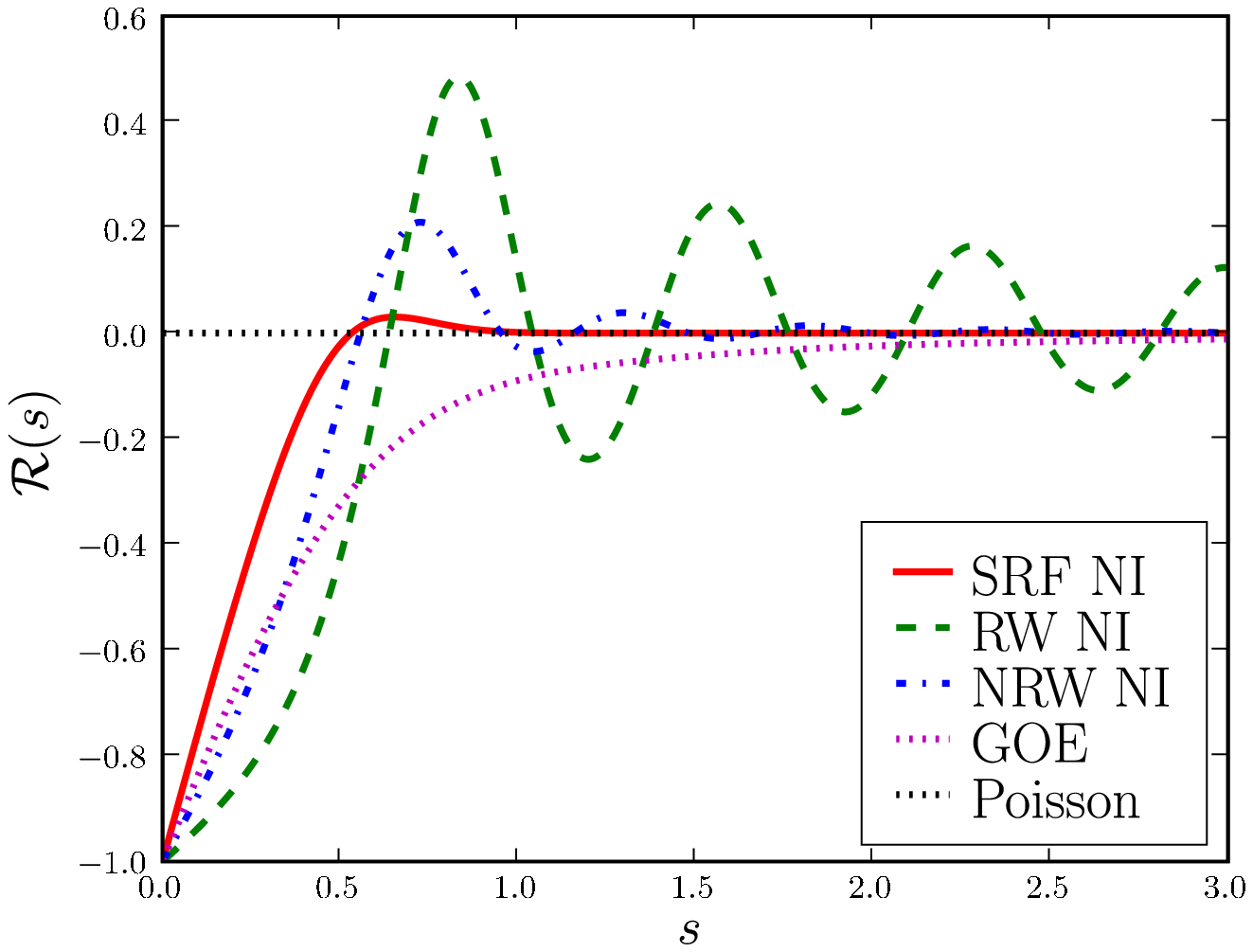}
  }
  \hspace{0.5cm}
  \subfloat[Form Factor]{\label{cap:srf_correl:ffact}
    \includegraphics[clip,scale=0.5]{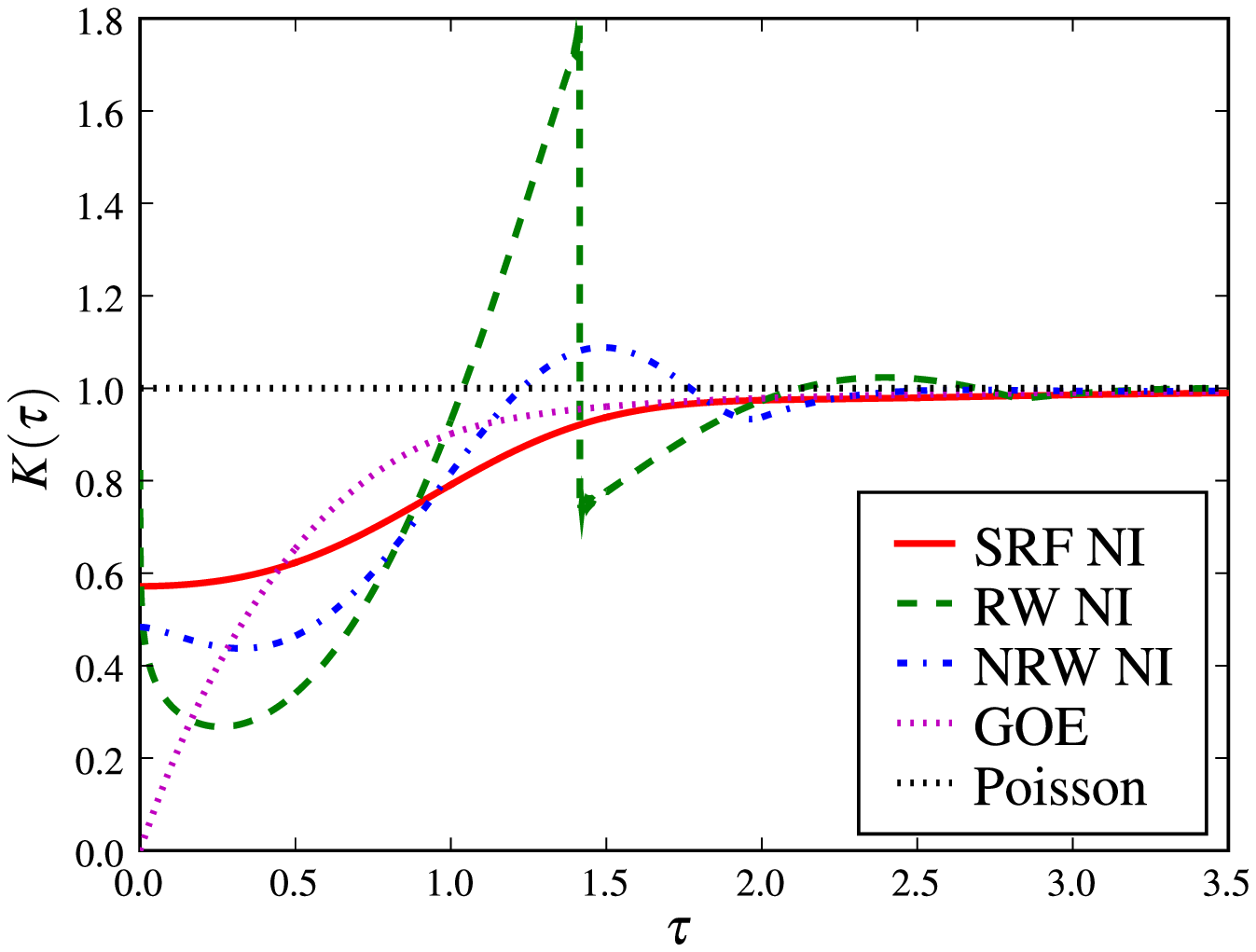}
  }
  \caption{\label{cap:srf_correl}SRF nodal intersections compared to other fields.}
\end{figure}

The expansion for small $s$ is:
\[
\mathcal{R} \sim -1 + \frac{\pi^2}{4}s - \frac{\pi^4}{48}s^3
       + \frac{\sqrt{3}\pi^4}{54}s^4 +\Or\!\!\left(s^5\right) ,
\]
and for large $s$, it is:
\[
\mathcal{R}\sim e^{-{(\pi s)}^2}\left[\frac{1}{2}{(\pi s)}^4 - 2{(\pi s)}^2 + 1\right]
      +\Or\!\!\left(e^{-2{(\pi s)}^2} s^8\right) .
\]
As can be seen in~\fref{cap:srf_correl:corrs}, the slope at 0 is higher than
that of the GOE, and the correlations decay to 0 faster than all other 
distributions considered here. 
Thus, this field has smaller repulsion and it may be considered the closest to
the random Poisson distribution.

The number variance and form factor also reflect this property. 
Denoting
\begin{eqnarray*}
\mathcal{M}^0_{\mathrm{SRF}} = \int_0^\infty \mathcal{R}(s)\rmd s \sim -0.2141 \\
\mathcal{M}^1_{\mathrm{SRF}} = \int_0^\infty s \mathcal{R}(s)\rmd s \sim -0.02902 \\
\mathcal{M}^2_{\mathrm{SRF}} = \int_0^\infty s^2 \mathcal{R}(s)\rmd s \sim -0.00434
\end{eqnarray*}

We find that the asymptotic number variance is linear with coefficient
$(1+2\mathcal{M}^{0}_{\mathrm{SRF}})\sim 0.5718$, which is larger than the
normally derived random waves (thus closer to Poisson).

The form factor has a smooth form (because~\eref{eq:short_covar_line} is not
oscillatory). The expansion for small $\tau$ is:
\[
K(\tau) \sim (1+2\mathcal{M}^{0}_{\mathrm{SRF}}) 
        - 4\pi^2\mathcal{M}^{2}_{\mathrm{SRF}} \tau^2
\]
We get a relatively large $K(0)\sim 0.5718$, and a quick approach to
the limiting value of 1 (which is consistent with the notion that this
is closer to the Poisson distribution).

\subsection{Chaotic billiards}

Eigenfunctions of chaotic billiards provide an ensemble of functions with
complex behaviour. In the semiclassical limit, they are well modelled by
Gaussian random wave ensembles (the monochromatic random wave ensemble is used
to model the eigenfunctions in regions which are far enough
from the boundary, and various boundary modified ensembles model the field
close to the boundary).

The statistics considered in this paper were evaluated numerically for 
eigenfunctions of desymmetrized Bunimovich stadium ($0.5\times 1$ rectangle
joined with a quarter of a circle of radius $1$) and desymmetrized Sinai 
billiard ($1.2\times 1$ rectangle, with a quarter of a circle of radius $0.5$
cut out of one of the corners)---both with Dirichlet boundary conditions.
For the Stadium, 1500 eigenfunctions were taken, with wave numbers 
ranging from $k=110$ to $k=165$. For the Sinai billiard, we used 10000 wave 
functions, with wave numbers from 350 to 500.

\begin{figure}[htbp]
  \centering
  \subfloat[Stadium billiard]{\label{cap:stad_line}
     \includegraphics[clip,scale=0.6]{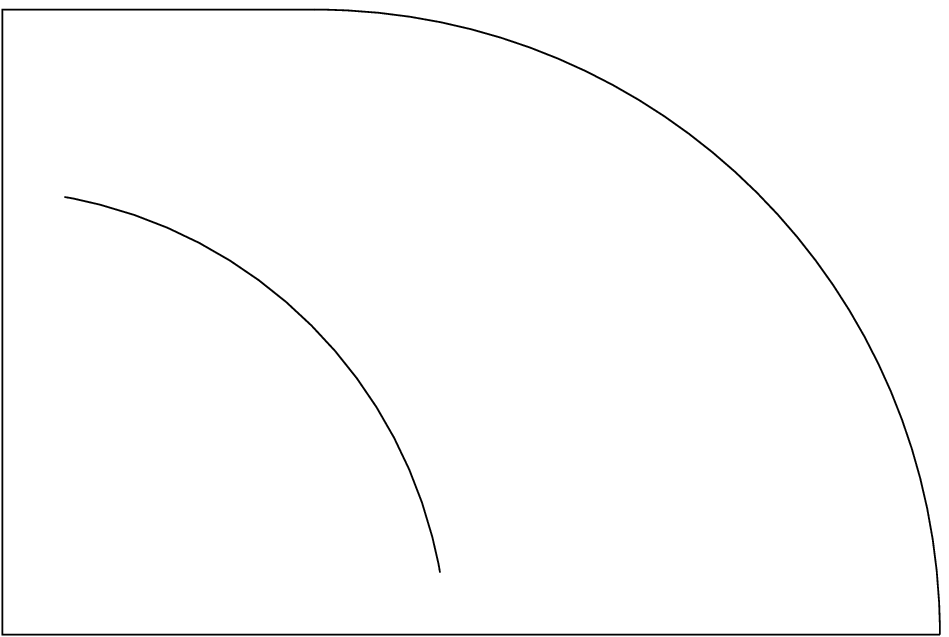}
  }
  \hspace{0.5cm}
  \subfloat[Sinai billiard]{\label{cap:sinai_line}
    \includegraphics[clip,scale=0.6]{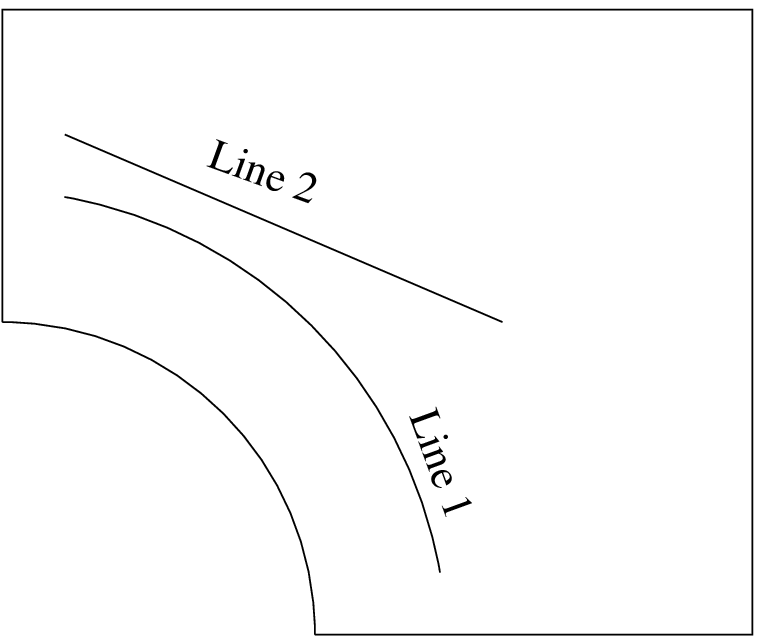}
  }
  \caption{\label{cap:intlines}Internal reference curves in the quantum 
billiards}
\end{figure}

\begin{figure}[htbp]
  \centering
  \subfloat[NN spacing]{\label{cap:bil_in:nn}
     \includegraphics[clip,scale=0.5]{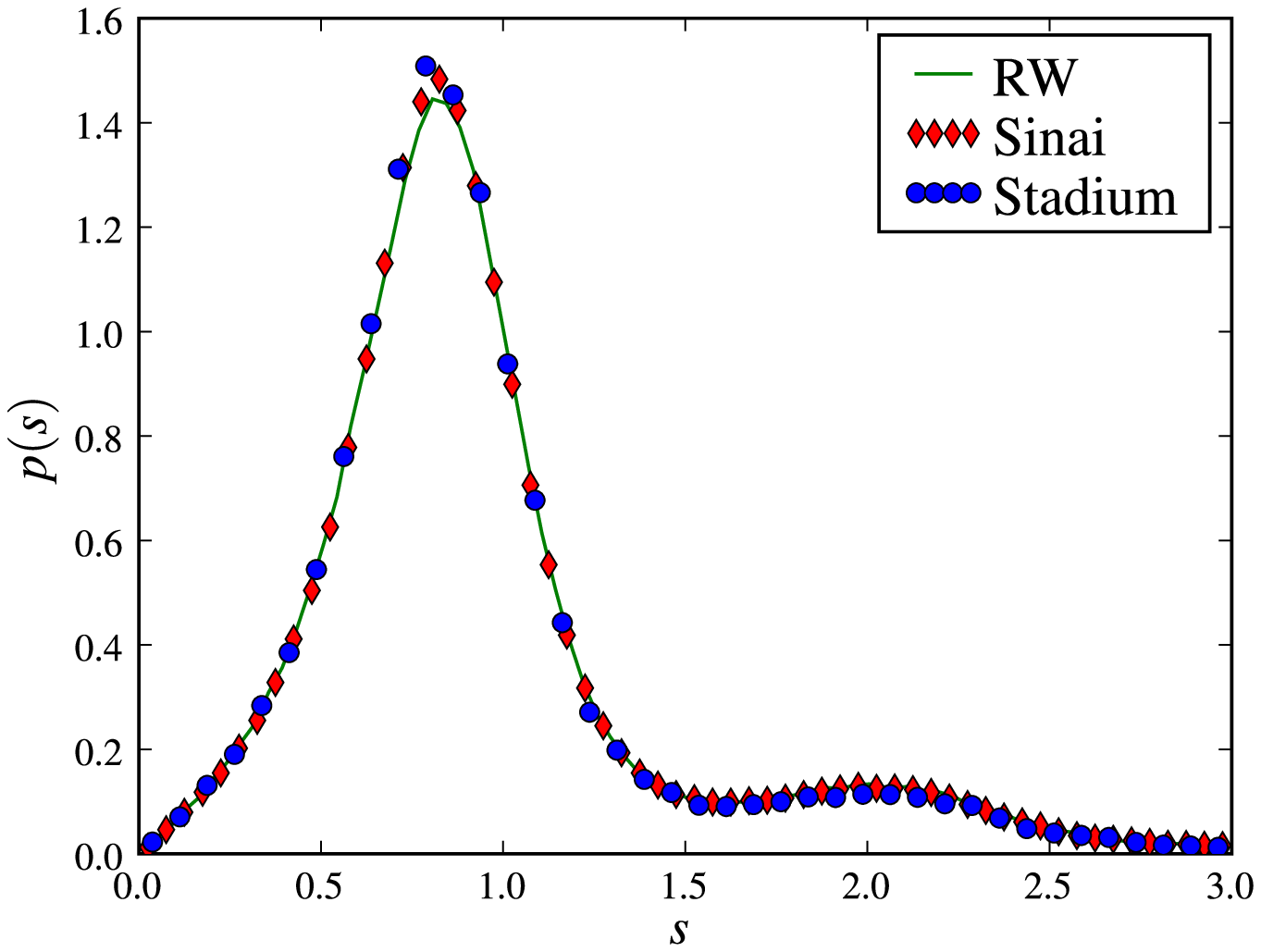}
  }
  \hspace{0.5cm}
  \subfloat[Form factor]{\label{cap:bil_in:ff}
    \includegraphics[clip,scale=0.5]{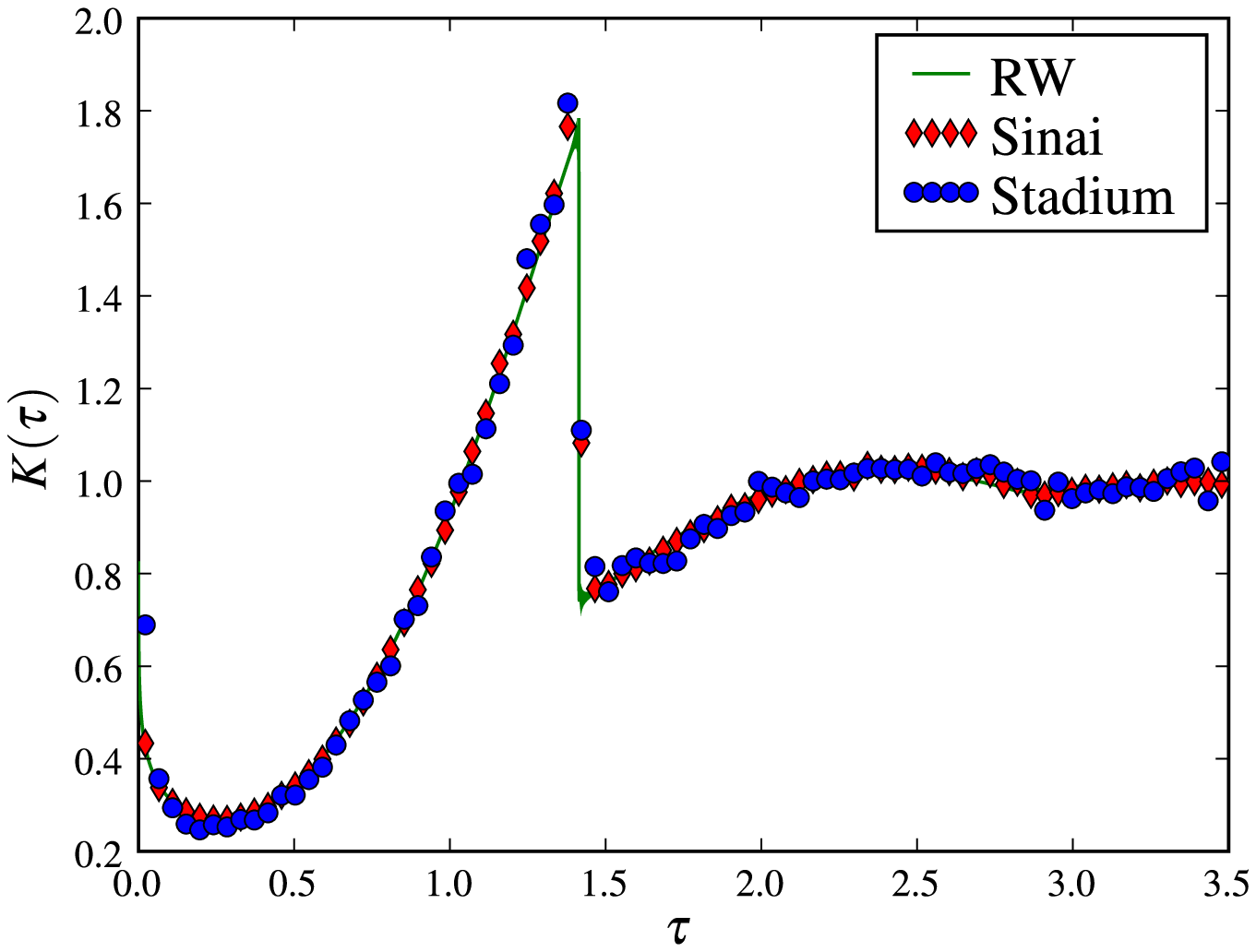}
  }
  \caption{\label{cap:bil_in}Billiards: NI with internal line.}
\end{figure}

Fixed reference curves were chosen in the interior of the billiards, as shown
in~\fref{cap:intlines}. For each wave function, the sequence of intersections
of the nodal lines with these curves was calculated, and normalized to unit
average spacing according to the corresponding wavenumber.
As shown in~\fref{cap:bil_in}, the nearest neighbour distribution agrees very
well with the predictions of the random waves model.
The form factor is also very close to the expected curve.

\begin{figure}[htbp]
  \centering
  \includegraphics[clip,scale=0.7]{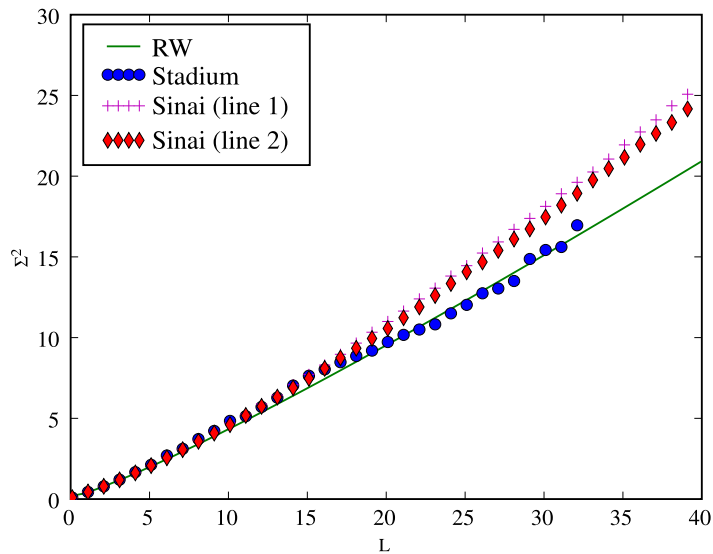}
  \caption{\label{cap:numvar_bil}Number variance of NI with segments inside
           chaotic billiards.}
\end{figure}

The number variance of nodal intersections on segments of the reference curves
is plotted in~\fref{cap:numvar_bil}. For short segments, the results are close
to the predictions of the random waves model, but deviate considerably from this
model as we move to segments of large normalized length, especially in the case
of the Sinai billiard. This deviation happens when the (unscaled) length of the
segment is comparable to the billiard dimensions. On such scales, it seems 
likely that the geometrical details of the billiard would have an effect on the
statistical properties of the wave functions.

\section{Summary}

The analytic results obtained for the three Gaussian fields discussed in this
paper are summarized in~\tref{cap:comparetab}, and compared to their analogues
in the well known random matrix and Poisson ensembles. For short distances,
the normalized correlation function for all the distributions considered 
 (except of the Poisson distribution) approaches $-1$, so the leading term of 
$1+\mathcal{R}(s)$ is listed, for $s=\epsilon\ll 1$. Similarly, the number
variance for short intervals is asymptotically $L-L^2$ (again, except of the
Poisson case),
so the leading term of $\hat{R}_2 = \Sigma^2(L)-L+L^2=\mean[n(n-1)]$ is listed.
 (for $L=\epsilon\ll 1$, this is also twice the probability for finding two
points in the interval). For large distances, we list the leading terms of
$\Sigma^2$ and $\mathcal{R}$. For the form factor, we list the ``asymptotic
rigidity'' $K(0)$ (which is directly related to $\Sigma^2$ and $\bar{\Delta}$
in large intervals~\cite{mehta:rand_mat},\cite{bohigas:rmt}), the
singular frequencies where $K(\tau)$ or its derivatives diverge and the type of
divergence (number of the first diverging derivative) in the first non zero 
singular frequency.

\begin{table}[htbp]
  \caption{\label{cap:comparetab}Comparison of NI statistics.}
  \lineup{}
  \begin{tabular*}{\textwidth}
    {@{}l @{\extracolsep{6pt plus 12pt}}l@{}l 
          @{}l@{}l
          @{\extracolsep{6pt plus 12pt}}l@{}l}
\br
  &  \centre{2}{Correlation} & \centre{2}{Form Factor} & 
     \centre{2}{Number Variance} \\
\ns
  {Process}  & \crule{2}  & \crule{2}  & \crule{2} \\
     & $1+\mathcal{R}(\epsilon)$ & $\mathcal{R}(s\gg 1)$ & 
                $K(0)$ & {Divergences$^{\dagger}$} &
 $\hat{R}_2(\epsilon)$ & $\Sigma^2(L\gg 1)$ \\
\mr
{Poisson} &
                         $1$ & $\m 0$ &
                        $1$  & \centre{1}{---\phantom{\{\}}}  &
$\phantom{0.000}\epsilon^2$  & $\phantom{0.000}L$ \\
\ms
{SRF} & 
             $2.467\epsilon$ & $\m 48.7s^4 \rme^{-{(\pi s)}^2}$ &
                     $0.572$ & \centre{1}{---\phantom{\{\}}} &
           $0.822\epsilon^3$ & $0.572L$         \\
{NRW} & 
             $1.234\epsilon$ & $\m 0.023s^{-3}$ &
                     $0.484$ & ${}^{(3)}\{2,4,\ldots\}$&
           $0.411\epsilon^3$ & $0.484L$         \\
{RW} & 
             $0.617\epsilon$ & $\m 0.036s^{-1}$ &  
                    $\infty$ & $^{(1)}\{\sqrt{2},2\sqrt{2},\ldots\}$&
           $0.206\epsilon^3$ & $0.072 L\log L$  \\
\ms
{GOE} & 
             $1.645\epsilon$ & $-0.101s^{-2}$   &
                         $0$ & ${}^{(4)}\{1\}$  &
           $0.548\epsilon^3$ & $0.203\log L$    \\
{GUE} &
           $3.290\epsilon^2$ & $-0.051s^{-2}$   &
                         $0$ & ${}^{(2)}\{1\}$  &
           $0.548\epsilon^4$ & $0.101\log L$    \\
{GSE} &
           $11.55\epsilon^4$ & $\m 0.25s^{-1}\cos(\omega s)$ &
                         $0$ & ${}^{(0)}\{1,2\}$&
           $0.770\epsilon^6$ & $0.051\log L$    \\
\br
\end{tabular*}
\noindent $^\dagger$ Nonzero values of $\tau$, for which $K(\tau)$ or one of
its derivatives diverge. The number in parentheses specifies the diverging 
derivative at the first of these points.
\end{table}

We have seen that
\begin{itemize}
\item Nodal intersections (NI) with a reference curve can be used as a 
statistical means to obtain insight into the properties of complex networks of
lines. The NI statistics of the random wave (RW) model display characteristic
behaviour, which is quite different from the statistics of other point processes.
Eigenfunctions of chaotic billiards match these statistics well, with some 
differences on scales comparable to the dimensions of the billiard.
\item For the RW and some other Gaussian models, the two point correlations and
other statistics related to it can be derived by analytical means.
\item The nearest neighbour distribution is harder to obtain (since it depends
on $n$-point correlations with $n\ge 3$). However, numerical evidence suggest
that the frequency of oscillations is related to the dominant frequencies
calculated for the corresponding two point correlation functions.
\end{itemize}

\FloatBarrier
\ack

This research was initiated as a result of a conversation with Percy Deift,
whose interest and help accompanied us throughout the work. Thanks Percy!
We would like to acknowledge Sven Gnutzmann for valuable discussion
and for the Stadium billiard data,
Zeev Rudnick and Mark Dennis for important references. We thank Holger Schanz 
for the Sinai billiard data, and Yehonatan Elon for careful reading of the text
and useful comments.

The work was supported by the Minerva Center for non-linear
Physics and the Einstein (Minerva) Center at the Weizmann Institute,
and by grants from the GIF (grant I--808--228.14/2003), and EPSRC
 (grant GR/T06872/01.) and a Minerva grant.

\appendix

\section{Derivation of the zeros correlation from amplitude statistics}
\label{sec:deriv_corr}

In~\sref{sub:inters_stats}, the correlation of the zeros of a Gaussian random
function on a curve was expressed in terms of the Fourier 
integral~\eref{eq:corr_fourdef}:
\begin{equation*}
\fl
\langle \rho \rho' \rangle = 
  \left\langle
    \frac{1}{4\pi^{4}}
    \int\frac{\rmd\xi\,\rmd\eta\,\rmd\xi'\,\rmd\eta'}{\eta^{2}\eta'^{2}}
       \rme^{\rmi\xi f}   (1-\rme^{\rmi\eta \dot{f}})
       \rme^{\rmi\xi' f'} (1-\rme^{\rmi\eta' \dot{f}'})
  \right\rangle
\end{equation*}

We note that the integrand contains four terms of the form 
$\exp(i\bi{v}_i\cdot\bi{f})$, with
\begin{eqnarray*}
\fl \bi{v}_i \in \{ (\xi,\xi',0,0),(\xi,\xi',\eta,0),
                (\xi,\xi',0,\eta'), (\xi,\xi',\eta,\eta') \}, \\
\fl \mathrm{and}\quad \bi{f}=(f,f',\dot{f},\dot{f}') .
\end{eqnarray*}
If we use~\eref{eq:expmulti} to evaluate the statistical average, these terms
are replaced by $\exp\!\!\left(-\frac{1}{2}\bi{v_i}^T M \bi{v_i} \right)$,
where $M$ is the covariance matrix of $\bi{f}$, given in~\eref{eq:amp_corrmat}.
To continue, we first integrate the $\xi$ and $\xi'$ coordinates. To separate
them out, we use a simple block matrix identity.

For any pair of column vectors $\bi{x},\bi{y} \in \mathbb{R}^n$ and three 
$n\times n$ matrices $A,B,C$, where $A$ and $B$ are symmetric and $A$ is 
non-singular, the following identity can be simply derived:
\begin{equation}
\fl
(\bi{x}^{T}\bi{y}^{T})
\left(\begin{array}{cc}
       A & C\\
       C^{T} & B
      \end{array}\right)
\left(\begin{array}{c}
       \bi{x}\\
       \bi{y}
      \end{array}\right) = 
    (\bi{x}^{T}+\tilde{\bi{y}}^T)A(\bi{x}+\tilde{\bi{y}})
    +\bi{y}^T\tilde{B}\bi{y}
\label{eq:mat_trick}
\end{equation}
where $\tilde{B}=B - C^T A^{-1} C$ and $\tilde{\bi{y}}=A^{-1} C \bi{y}$.

In addition, using the same notation, one can also deduce the determinant
identity
\begin{equation}
\det \left(\begin{array}{cc}
             A & C\\
             C^{T} & B\end{array}\right) =
\det A \det \tilde{B},
\label{eq:det_trick}
\end{equation}
which we shall use below.

Applying~\eref{eq:mat_trick} to the four terms, with 
${\bi{v}_i}^T = (\bi{x}^T,{\bi{y}_i}^T)$ (i.e.\ $\bi{x}^T=(\xi,\xi')$ and
${\bi{y}_i}^T \in \{(0,0),(\eta,0),(0,\eta'),(\eta,\eta')\}$), the $\bi{x}$
dependent integration for each of the four terms reduces to the usual Gaussian
form:
\[
\int \rmd\xi \rmd \xi' 
   \rme^{-\frac{1}{2}(\bi{x}^T+{\tilde{\bi{y}}_i}^T)A(\bi{x}+\tilde{\bi{y}}_i)}
  = \frac{2\pi}{\sqrt{\det A}}
\]
This does not depend on $\bi{y}$ and therefore contributes a constant 
multiplicative factor. The remaining integral assumes the following form:
\[ \fl
\int\!\!\!\int_{-\infty}^{\infty}\frac{\rmd\eta\rmd\eta'}{\eta^2\eta'^2}
    \left(1-\rme^{-\frac{1}{2}a\eta^2}-\rme^{-\frac{1}{2}b\eta'^2}
          +\rme^{-\frac{1}{2} {\bi{y}_2}^T \tilde{B} \bi{y}_2}
    \right)
\]
 (using $a$,$b$ and $c$ for the matrix elements of $\tilde{B}$).

This can be rewritten as $I_1+I_2$, where
\[ \fl
I_1 = \int\!\!\!\int_{-\infty}^{\infty}
   \frac{\rmd\eta\rmd\eta'}{\eta^2\eta'^2}
     (1-\rme^{-\frac{1}{2}a\eta^2})(1-\rme^{-\frac{1}{2}b\eta'^2})
    =2\pi\sqrt{ab}
\]
and 
\[ \fl
I_2 = \int\!\!\!\int_{-\infty}^{\infty}
   \frac{\rmd\eta\rmd\eta'}{\eta^2\eta'^2}
     \rme^{-\frac{1}{2}(a\eta^2+b\eta'^2)}(\rme^{-c\eta\eta'}-1)
\]

$I_2$ is not well defined, but taking the Cauchy Principal Value
for the integrals, anti-symmetric integrands do not contribute. After 
symmetrization and change of integration variables, we get:
\[ \fl
I_2 = \sqrt{ab}\int\!\!\!\int_{-\infty}^{\infty}
   \frac{\rmd x\rmd y}{x^{2}y^{2}}
   \rme^{-\frac{1}{2}(x^2+y^2)} \left[ \cosh(\hat{c}xy)-1 \right]
\]
with $\hat{c}\equiv c/\sqrt{ab}$. One way to solve this integral is to derive
the expression by $\hat{c}$ twice, and integrate back after the $x$ and $y$
integrations. Another way is to express 
$f(xy)\equiv {(xy)}^{-2}[\cosh(\hat{c}xy)-1]$ as a power series in ${(xy)}^2$,
and integrate term by term (each term becomes a square of a simple Gaussian
moment). Either way, for any $|\hat{c}|<1$, we get
\[ \fl
I_2 = 2\pi \sqrt{ab}\left[-1+\sqrt{1-\hat{c}^{2}}
                          +\hat{c}\arcsin\hat{c}\right],
\]
To show that the convergence condition $|\hat{c}|<1$ holds, we note that
\[
|\hat{c}|<1 \Leftrightarrow 
\det\tilde{B} = ab - c^2 > 0.
\]
However, from~\eref{eq:det_trick} we have $\det M=\det A \det\tilde{B}$. Since 
both $A$ and $M$ are covariance matrices, they are positive definite, and 
their determinant must be positive for any non-degenerate case. This can only
be consistent if $\det \tilde{B}$ is positive too.

Inserting $I_1$ and $I_2$ into the expression for the normalized correlation we
find~\eref{eq:correl}
\begin{equation*} \fl
\mathcal{R}(t,t')= 
  \frac{c_0}{c_2} \frac{a}{{(\det A)}^{3/2}}
  \left(\sqrt{1-\hat{c}^{2}}+\hat{c}\arcsin(\hat{c})\right) - 1
\end{equation*}

\section{Calculating the amplitude correlations for the normal derivative of
a Gaussian field}
\label{sec:curv_ndrv_ampcorr}

When the investigated field is translationally invariant and isotropic, the
correlation functions $C_0$, $C_1$ and $C_2$ for the ``normally derived''
function $g(t)=\bi{n}(t)\cdot \bi{\triangledown}\psi(\bi{r}(t))$ can be
calculated in terms of the curve parameters and reduced correlation function
$G(d)$. The derivation is straightforward but tedious, and follows the same 
mechanism used in~\sref{sub:restrict} and~\sref{sub:bm_restrict}. In this 
section we recite the main steps and results.

As in previous sections, we start by using linearity to move the derivatives
out of the statistical average, getting~\eref{eq:curv_ndrv_c0} for the first
correlation function $C_0$. This expression involves second order partial 
derivatives of the basic correlation $\langle\psi \psi'\rangle$, so the 
expressions for $C_1$ and $C_2$ involve third and fourth order partial 
derivatives correspondingly.

Using shorthand notation where tagged parameters
refer to values at $t'$, $\partial_i\equiv \partial/ \partial r_i$ and
${\partial'}_i\equiv \partial/ \partial {r'}_i$, we have
\begin{eqnarray}
C_0 = & \sum_{i,j} n_i {n'}_j \partial_i {\partial'}_j G(|\bi{r}-\bi{r}'|) 
  \nonumber \\
C_1 = & \sum_{i,j} {\partial'}_t n_i {n'}_j \partial_i {\partial'}_j G
  \nonumber \\
C_2 = & \sum_{i,j}\partial_t{\partial'}_t n_i{n'}_j \partial_i {\partial'}_j G
\label{eq:ndrv_coeffs_curv_def}
\end{eqnarray}
The derivatives of $\bi{n}$ in these expressions can be evaluated using Frenet 
formula $\partial_t \bi{n} = -\kappa \dot{\bi{r}}$ (where $\kappa$ is the
curvature at $t$), and the partial derivatives of $G$ can be expressed in
terms of derivatives with respect to $d$ ($G'$,$G''$,$G^{(3)}$,$G^{(4)}$) and
the distance vector $\bi{d}$.

First, we calculate the partial derivatives of $G$ using the following rules 
for derivatives of the distance $d \equiv |\bi{d}|$ and direction vector 
$\hat{\bi{d}} \equiv \bi{d}/d$ (where $\bi{d} = \bi{r}-\bi{r'}$):
\begin{eqnarray*}
\partial_i d = -{\partial'}_i d = \hat{d}_i \\
\partial_i \hat{d}_j = -{\partial'}_i \hat{d}_j 
   = \frac{1}{d} \left(\delta_{ij} - \hat{d}_i \hat{d}_j\right)
\end{eqnarray*}

To express the results in a manifestly symmetric way, we define 
$d^{[k]}_{i_1,i_2,\ldots,i_N}$, the ``$N$ dimensional symmetric direction 
tensor of degeneracy $k$''
as the symmetrized product of $k$ Kronecker deltas and $N-2k$ direction
vectors $\hat{\bi{d}}$. Specifically:
\begin{eqnarray*}
    d^{[0]}_{ij}  =& \hat{d}_i\hat{d}_j ,\quad d^{[1]}_{ij}=\delta_{ij} \\
    d^{[1]}_{ijk} =& \delta_{ij}\hat{d}_k + \delta_{ik}\hat{d}_j 
                  + \delta_{jk}\hat{d}_i\\
    d^{[1]}_{ijkl} =
            & \delta_{ij}\hat{d}_k\hat{d}_l + \delta_{ik}\hat{d}_j\hat{d}_l + 
              \delta_{il}\hat{d}_j\hat{d}_k + \\
            & \delta_{jk}\hat{d}_i\hat{d}_l + 
              \delta_{jl}\hat{d}_i\hat{d}_k + \delta_{kl}\hat{d}_i\hat{d}_j \\
    d^{[2]}_{ijkl} =& 
              \delta_{ij}\delta_{kl} + \delta_{ik}\delta_{jl} + 
              \delta_{il}\delta_{jk}
\end{eqnarray*}

We also introduce the notation $G^{[n]}$ for the specific combinations of
derivatives of $G$ up to order $n$, which satisfy the following recurrence 
relations\footnote{
These relations are almost identical to the Bessel recurrence relations. Hence,
for the case where $G(d)=J_0(kd)$, we get $G^{[n]}={(-k)}^n J_n(kd)$
}:
\begin{eqnarray*}
G^{[0]}(d) = G(d) \\
G^{[n+1]}(d) = {G^{[n]}}'(d) - \frac{n}{d}G^{[n]}(d)
\end{eqnarray*}

Specifically, we have:
\begin{eqnarray*}
G^{[1]} = G' \\
G^{[2]} = G''-\frac{1}{d}G' \\
G^{[3]} = G^{(3)}-\frac{3}{d}G''+\frac{3}{d^2}G'\\
G^{[4]} = G^{(4)}-\frac{6}{d}G^{(3)}+\frac{15}{d^2}G''-\frac{15}{d^3}G'
\end{eqnarray*}

With these notations, the partial derivatives of $G$ are given by:
\begin{eqnarray*}
\partial_i G = G^{[1]} d_i^{[0]} \\
\partial_i \partial_j G =
   G^{[2]}d_{ij}^{[0]} + \frac{1}{d}G^{[1]} d_{ij}^{[1]} \\
\partial_i\partial_j\partial_k G =
   G^{[3]}d_{ijk}^{[0]} + \frac{1}{d}G^{[2]}d_{ijk}^{[1]} \\
\partial_i\partial_j\partial_k\partial_l G =
   G^{[4]}d_{ijkl}^{[0]} + \frac{1}{d}G^{[3]}d_{ijkl}^{[1]}
   + \frac{1}{d^2}G^{[2]}d_{ijkl}^{[2]}
\end{eqnarray*}

Substituting these in~\eref{eq:ndrv_coeffs_curv_def}, we get
\begin{eqnarray}
C_0 & = \sum_{i,j} 
           -n_i {n'}_j(G^{[2]}d_{ij}^{[0]}
           +\frac{1}{d}G^{[1]}d_{ij}^{[1]}) \nonumber\\
    & =-(\bi{n}\cdot\hat{\bi{d}})(\bi{n}'\cdot\hat{\bi{d}})G^{[2]}
       -(\bi{n}\cdot\bi{n}')\frac{1}{d}G^{[1]}
\label{eq:ndrv_c0_curv}
\end{eqnarray}
which is equivalent to~\eref{eq:curv_ndrv_c0}. Similarly,
\begin{eqnarray}
\fl C_1&=&\sum_{i,j}
          \kappa'n_i {\dot{r}'}_{j} \partial_i\partial_j G
       +\sum_{i,j,k}
          n_i {n'}_j {\dot{r}'}_k \partial_i\partial_j\partial_k G \nonumber \\
\fl &=&\kappa'\left[(\bi{n}\hat{\bi{d}})(\dot{\bi{r}}'\hat{\bi{d}})G^{[2]} +
                (\bi{n}\dot{\bi{r}}')\frac{1}{d}G^{[1]}\right] \nonumber\\
\fl & &+ (\bi{n}\hat{\bi{d}})(\bi{n}'\hat{\bi{d}})
            (\dot{\bi{r}}'\hat{\bi{d}})G^{[3]}
       + \frac{1}{d}\left[ (\bi{n}\bi{n}')(\dot{\bi{r}}'\hat{\bi{d}})
                           +(\bi{n}\dot{\bi{r}}')(\bi{n}'\hat{\bi{d}})
                    \right] G^{[2]}
\label{eq:ndrv_c1_curv}
\end{eqnarray}

and, defining $\tilde{\bi{d}}=\kappa'\bi{n} - \kappa\bi{n}'$, we get

\begin{eqnarray}
\fl C_2 &=& 
   -\kappa\kappa' \dot{r}_i {\dot{r}'}_j \partial_{ij}G 
   + \tilde{d}_i \dot{r}_j{\dot{r}'}_k \partial_{ijk}G 
   + n_i {n'}_j \dot{r}_k {\dot{r}'}_l \partial_{ijkl}G \nonumber\\
\fl     &=&   
   - \kappa \kappa'
     \left[(\dot{\bi{r}}\hat{\bi{d}})(\dot{\bi{r}}'\hat{\bi{d}})G^{[2]} +
           (\dot{\bi{r}}\dot{\bi{r}}')\frac{1}{d}G^{[1]}\right] \nonumber\\
   \fl  & &   
   +(\tilde{\bi{d}}\hat{\bi{d}})(\dot{\bi{r}}\hat{\bi{d}})
    (\dot{\bi{r}}'\hat{\bi{d}})G^{[3]} \nonumber\\
      \fl  & &
      \qquad +\left[ (\tilde{\bi{d}}\dot{\bi{r}}) (\dot{\bi{r}}'\hat{\bi{d}})
                    +(\tilde{\bi{d}}\dot{\bi{r}}') (\dot{\bi{r}}\hat{\bi{d}})
                    +(\tilde{\bi{d}}\hat{\bi{d}}) (\dot{\bi{r}}\dot{\bi{r}}')
              \right] \frac{1}{d}G^{[2]} \nonumber\\
   \fl  & &   
   +(\bi{n}\hat{\bi{d}})(\bi{n}'\hat{\bi{d}})
    (\dot{\bi{r}}\hat{\bi{d}})(\dot{\bi{r}}'\hat{\bi{d}})G^{[4]}\nonumber\\
      \fl  & &   
      \qquad +\left[ (\bi{n}\hat{\bi{d}})(\bi{n}'\hat{\bi{d}})
                     (\dot{\bi{r}}\dot{\bi{r}}')
                    +(\bi{n}\hat{\bi{d}})(\dot{\bi{r}}'\hat{\bi{d}})
                     (\dot{\bi{r}}\bi{n}') \right. \nonumber\\
                \fl  & & \qquad \quad
                \left.
                    +(\bi{n}'\hat{\bi{d}})(\dot{\bi{r}}\hat{\bi{d}})
                     (\bi{n}\dot{\bi{r}}')
                    +(\dot{\bi{r}}\hat{\bi{d}})(\dot{\bi{r}}'\hat{\bi{d}})
                     (\bi{n}\bi{n}')
              \right] \frac{1}{d}G^{[3]} \nonumber \\
       \fl & &    
       \qquad +\left[ (\bi{n}\bi{n}')(\dot{\bi{r}}\dot{\bi{r}}')
                     +(\bi{n}\dot{\bi{r}}')(\dot{\bi{r}}\bi{n}')
               \right] \frac{1}{d^2}G^{[2]}
\label{eq:ndrv_c2_curv}
\end{eqnarray}

For the case where the curve is a circle, 
equations~\eref{eq:ndrv_c0_curv}--\eref{eq:ndrv_c2_curv} can be much simplified.
Denoting $\alpha=\frac{1}{2}\kappa(t-t')$ as
in~\sref{sec:mrw}, the following identities hold:
\begin{eqnarray*}
\fl \kappa = \kappa', \quad d = \frac{2}{\kappa}\sin\alpha,
  \quad \tilde{\bi{d}} = -2 \kappa\sin\alpha \hat{\bi{d}} \\
\fl \hat{\bi{d}}\cdot\dot{\bi{r}} = \hat{\bi{d}}\cdot\dot{\bi{r}}'
         = \cos(\alpha),
  \quad \hat{\bi{d}}\cdot\bi{n} = -\hat{\bi{d}}\cdot\bi{n}' = -\sin(\alpha) \\
\fl \bi{n}\cdot\bi{n}' = \dot{\bi{r}} \cdot \dot{\bi{r}}' = \cos(2\alpha),
  \quad \dot{\bi{r}}\cdot\bi{n}' = -\dot{\bi{r}}' \cdot \bi{n} = \sin(2\alpha)
\end{eqnarray*}

Substituting these in~\eref{eq:ndrv_c0_curv}--\eref{eq:ndrv_c2_curv}, and
expressing the trigonometric coefficients in terms of $c\equiv\cos\alpha$,
the correlations become:
\begin{eqnarray}
\fl C_0 =& (1-c^2)G^{[2]}+(1-2c^2)G^{[1]}\frac{1}{d} \nonumber\\
\fl C_1 =& (c^3-c)G^{[3]}+(6c^3-5c)G^{[2]}\frac{1}{d}
          +(4c^3-4c)G^{[1]}\frac{1}{d^2} \nonumber\\
\fl C_2 =& (c^4-c^2)G^{[4]}+(12c^4-12c^2+1)G^{[3]}\frac{1}{d} \nonumber\\
     \fl &+(28c^4-32c^2+5)G^{[2]}\frac{1}{d^2}
          +(8c^4-12c^2+4)G^{[1]}\frac{1}{d^3}
\label{eq:ndrv_curv_sym}
\end{eqnarray}

or, in terms of the actual derivatives $G^{(n)}$:
\begin{eqnarray*}
\fl C_0 =& (1-c^2)G''+(-c^2)G'\frac{1}{d} \\
\fl C_1 =& (c^3-c)G^{(3)}+(3c^3-2c)G''\frac{1}{d}
          +(c^3-2c)G'\frac{1}{d^2} \\
\fl C_2 =& (c^4-c^2)G^{(4)}+(6c^4-6c^2+1)G^{(3)}\frac{1}{d} \\
     \fl &+(7c^4-11c^2+2)G''\frac{1}{d^2}
          +(c^4-c^2+2)G'\frac{1}{d^3}
\end{eqnarray*}
which is equivalent to~\eref{eq:circ_ndrv_c0}--\eref{eq:circ_ndrv_c2}.

\section{Using asymptotic series to expand the integral transforms}
\label{sec:asymp_integs}

In~\sref{sub:var_power} and other sections, we discuss the number variance and
form factor of some distributions. These statistics can be expressed
as an integral involving the normalized correlation (\eref{eq:numvar} 
and~\eref{eq:formfact}), for which we have an explicit expression. The 
expression for the correlation function (e.g.~\eref{eq:rw_covar_line} 
or~\eref{eq:ndrv_covar_line}) is usually very complicated, which makes the
resulting integral hard to solve directly.
However, given the asymptotic expansion of the normalized correlation, we can
calculate the asymptotic expansion of the integral up to a small number of
constants, which can be calculated numerically. The outline of this method is
described here.

When calculating the number variance for large $L$, we encounter integrals of
the form
\begin{equation}
I_f(L) = \int_0^L f(x)\rmd x
\label{eq:integ_tol}
\end{equation}
where $f(x)= x^k \mathcal{R}(x)$, and $\mathcal{R}(x)$ is the normalized
correlation, for which the asymptotic expansion is known. We split the
asymptotic expansion of $f$ into two parts:
\[
f(x) \sim \sum_{n=1}^{n_0}a_n(x) + \sum_{n=n_0+1}^\infty a_n(x) ,
\]
where $n_0$ is the number of terms $a_n$ such that $\int_L^\infty a_n(x)\rmd x$
diverges for every finite $L$ ($n_0$ may be 0, but we will assume it is 
finite. We will also assume that for $n>n_0$ the integrals converge uniformly
with some finite lower bound $L_0$).

Defining
\begin{equation}
D(x) \equiv \sum_{n=1}^{n_0}a_n(x),\quad C(x)\equiv f(x)-D(x),
\label{eq:asymp_split}
\end{equation}
we find that $C(x)$ generates an integral with known asymptotic expansion
\begin{equation}
\fl \int_L^\infty C(x)\rmd x \sim 
  \sum_{n=n_0+1}^{\infty} \int_L^\infty a_n(x)\rmd x
  \equiv - \sum_{n=n_0+1}^{\infty} b_n(L)
\label{eq:int_convpart}
\end{equation}

We now wish to add and subtract $D$ from the integrand of~\eref{eq:integ_tol}.
However, the interval $[0,L]$ might contain points which would cause the
integral of $D$ to diverge. For such cases, we split the interval at some point
$x_0$, which is larger than these problematic points, and do the subtraction
only in the second half $[x_0,L]$. We will assume that $x_0$ and $L$ are both
larger than $L_0$. In practice, for the functions used in this paper, there is
either no points of divergence (and $x_0$ is chosen to be 0), or a single point
of divergence at 0, in which case we choose $x_0=1$ (which happens to simplify
the resulting expression). Using~\eref{eq:asymp_split}, the integral becomes

\[
\fl I_f(L) = \int_0^{x_0} f(x)\rmd x + \int_{x_0}^L C(x) \rmd x + 
         \int_{x_0}^L D(x)\rmd x
\]

The third term in this expression consists of the ``diverging'' terms of $f$,
however the integral here is finite and its asymptotic expansion is easily
obtained from~\eref{eq:asymp_split}:
\begin{equation}
\fl \int_{x_0}^L D(x) \rmd x \sim
   \sum_{n=1}^{n_0} \int_{x_0}^{L} a_n(x)\rmd x
   \equiv \sum_{n=1}^{n_0} b_n(L)
\label{eq:int_divpart}
\end{equation}

To handle the remaining terms, we add and subtract~\eref{eq:int_convpart}
\begin{eqnarray*}
\fl I_f(L)&=& \int_0^{x_0} f(x)\rmd x + \int_{x_0}^\infty C(x) \rmd x  \\
\fl       & & + \int_{x_0}^L D(x)\rmd x - \int_{L}^\infty C(x) \rmd x
\end{eqnarray*}

In this expression, the first two terms are finite values independent of $L$,
and the last two have known asymptotic expansion. Using~\eref{eq:int_convpart}
and~\eref{eq:int_divpart}, we get the final result:
\begin{equation}
I_f(L) \sim \mathcal{M}_f + \sum_{n=1}^\infty b_n(L)
\end{equation}
where
\[
\fl \mathcal{M}_f \equiv 
   \int_0^{x_0} f(x)\rmd x + \int_{x_0}^\infty [f(x)-D(x)]\rmd x
\]

\newpage{}
\bibliographystyle{unsrt}
\bibliography{nodal}

\end{document}